\documentclass[%
 reprint,
superscriptaddress,
 amsmath,amssymb,
 aps,
]{revtex4-2}

\usepackage{graphicx}
\usepackage{dcolumn}
\usepackage{bm}
\usepackage[colorlinks=true,hyperfootnotes=true,breaklinks=true,citecolor=blue]{hyperref}
\usepackage{tabularray}
\usepackage{comment}

\begin{document}

\preprint{APS/123-QED}

\title{Quantum optical neural networks with programmable nonlinearities}

\author{E.\,A.\,Chernykh}
\affiliation {Quantum Technology Centre, M.\,V. Lomonosov Moscow State University, Leninskie Gory 1, Moscow, 119991, Russia}
\affiliation {Russian Quantum Center, Bolshoy Boulevard 30, building 1, Moscow, 121205, Russia}

\author{M.\,Yu.\,Saygin}
\email{saygin@physics.msu.ru}
\affiliation {Sber Quantum Technology Center, Kutuzovski prospect 32, Moscow, 121170, Russia}
\affiliation {Quantum Technology Centre, M.\,V. Lomonosov Moscow State University, Leninskie Gory 1, Moscow, 119991, Russia}
\affiliation{Laboratory of Quantum Engineering of Light, South Ural State University (SUSU), Prospekt Lenina 76, Russia, Chelyabinsk, 454080 Russia}

\author{G.\,I.\,Struchalin}
\affiliation {Quantum Technology Centre, M.\,V. Lomonosov Moscow State University, Leninskie Gory 1, Moscow, 119991, Russia}

\author{S.\,P.\,Kulik}
\affiliation {Quantum Technology Centre, M.\,V. Lomonosov Moscow State University, Leninskie Gory 1, Moscow, 119991, Russia}
\affiliation{Laboratory of Quantum Engineering of Light, South Ural State University (SUSU), Prospekt Lenina 76, Russia, Chelyabinsk, 454080 Russia}

\author{S.\,S.\,Straupe}
\affiliation {Sber Quantum Technology Center, Kutuzovski prospect 32, Moscow, 121170, Russia}
\affiliation {Russian Quantum Center, Bolshoy Boulevard 30, building 1, Moscow, 121205, Russia}
\affiliation {Quantum Technology Centre, M.\,V. Lomonosov Moscow State University, Leninskie Gory 1, Moscow, 119991, Russia}

\begin{abstract}
Parametrized quantum circuits are essential components of variational quantum algorithms. Until now, optical implementations of these circuits have relied solely on adjustable linear optical units. In this study, we demonstrate that using programmable nonlinearities, rather than linear optics, offers a more efficient method for constructing quantum optical circuits --- especially quantum neural networks. This approach significantly reduces the number of adjustable parameters  and the circuit depth needed to achieve high-fidelity operation. Specifically, we explored a quantum optical neural network (QONN) architecture composed of meshes of two-mode interferometers programmable by adjustable Kerr-like nonlinearities. We assessed the capabilities of our quantum optical neural network  architecture and compared its performance to previously studied architectures that use multimode linear optics units. Additionally, we suggest future research directions for improving programmable quantum optical circuits.

\end{abstract}

\maketitle


\section*{\label{sec:introduction}Introduction}

In recent years, quantum approaches to information processing, once confined to textbooks, have been successfully demonstrated in small-scale experiments across various physical platforms. While the ultimate goal remains the construction of a universal quantum computer --- a highly challenging task due to the stringent requirements for large-scale, high-quality physical resources --- current research focuses on extracting the maximum potential from the available noisy intermediate-scale quantum (NISQ) devices~\cite{PreskillNISQ}. In this context, variational quantum computing, particularly quantum machine learning methods, is considered one of the primary candidates to achieve quantum advantage for meaningful tasks~\cite{CerezoML,BiamonteML,PreskillML, NNs_Abbas2021}.

Variational quantum algorithms utilize parameterized quantum circuits, which are implemented on quantum processors and trained using classical optimization algorithms. On most common physical platforms, these circuits consist of layers of quantum gates, some of which are adjustable and operate within the qubit Hilbert space~\cite{CerezoVQE}. Since the majority of quantum platforms support standard gate sets by design, the states being manipulated remain tied to qubit encoding throughout the computation. However, while gate-based circuit descriptions are convenient and widely adopted, they may not be suitable for all quantum computing platforms.

The linear-optical platform exploits the non-classical properties of photons interfering in linear-optical circuits~\cite{ReckDesign, ClementsDesign,TimeBin_interferometer,RobustSaygin,FldzhyanDesign,Taballione2023modeuniversal,KondratyevLargeScale}, which form the foundation for both restricted~\cite{BOSON_SAMPLING_DV} and universal models of photonic quantum computing~\cite{KLM,MBQC,FBQC}. However, while linear-optical interference suffices for some algorithms, such as boson sampling~\cite{BOSON_SAMPLING_DV}, it is not enough to implement more universal algorithms directly. This has led to the development of alternative approaches tailored to photonic platforms, where entangling gates are probabilistically implemented through measurements~\cite{KLM,MBQC,FBQC}.

Our work draws inspiration from significant experimental progress made in developing non-linear elements that are sensitive to single-photon-level quantum states~\cite{RB_NPS_WISPERRING,Vapour_NS,SinglePhoton_NL_nature,Lodahl_arxiv_NS, Bulk_nonlinearity,TemporalTrapping}, especially those compatible with integrated photonic circuits~\cite{SinglePhoton_NL_nature,Lodahl_arxiv_NS,Bulk_nonlinearity,TemporalTrapping}. These elements can effectively introduce photon-photon interactions, which are lacking in linear optics~\cite{PhotonicQuantumTechnologies}. Additionally, similar approaches are being explored in the microwave domain using mature superconducting components, enjoying accessible strong nonlinearities required for the single-photon-level elements, in particular, deterministic nonlinear gates~\cite{Superconducting_NL}. These advances are motivating the exploration of quantum computing architectures that go beyond traditional linear-optical methods.

Quantum optical neural networks (QONNs) are a specific class of photonic circuits that alternate between multimode linear-optical evolution and single-site nonlinear elements, in order to manipulate photonic quantum states. Similar to classical neural networks, where training involves adjusting linear layers~\cite{DeepLearning}, previous research on QONNs has focused on training through adjustable  multimode linear optics~\cite{QuantumOpticalNNs, EwaniukQONNs}. In this paper, we introduce a novel approach to parameterized quantum optical circuits, specifically QONNs, which relies on programming single-site nonlinearities instead of programmable multimode linear optics. In particular, we study quantum optical circuits composed of nonlinear Mach-Zehnder interferometer (NMZI) meshes, where the only control parameters are the strengths of single-mode Kerr-like nonlinearities. We assess these circuits on a range of tasks, including generating entangled qubit states, performing Bell-state discrimination, and implementing variational quantum algorithms.

Our results show that QONNs based on this approach require significantly fewer adjustable parameters and  programmable optical element layers to achieve high-quality performance compared to those trained using linear-optics. This reduction translates into improved fidelities of QONNs. Thus, our findings suggest that variational quantum optical circuits utilizing programmable single-photon-sensitive nonlinearities offer substantial advantages.

The article is organized as follows. In Sec.~\ref{sec:qonn} we introduce the studied QONN that rely on programmable nonlinearities, rather than on programmable linear-optical interferometers. Sec.~\ref{sec:results} presents the QONNs performance analysis by several example tasks, namely, generation of specific photonic states, quantum operations and variational quantum eigensolver. We discuss the advantages of the QONN with variable nonlinearities and their possible realizations in Sec.~\ref{sec:discussion}.

\section{\label{sec:qonn}Quantum optical neural networks}

In this section, we introduce a QONN architecture, on which we demonstrate the proposed approach, its basic building blocks, and relevant parameters of QONNs, which quantify their hardware requirements.

\subsection{Variational core of the QONN}

\begin{figure*}[htbp]
        \centering
        \includegraphics[width=0.9\textwidth]{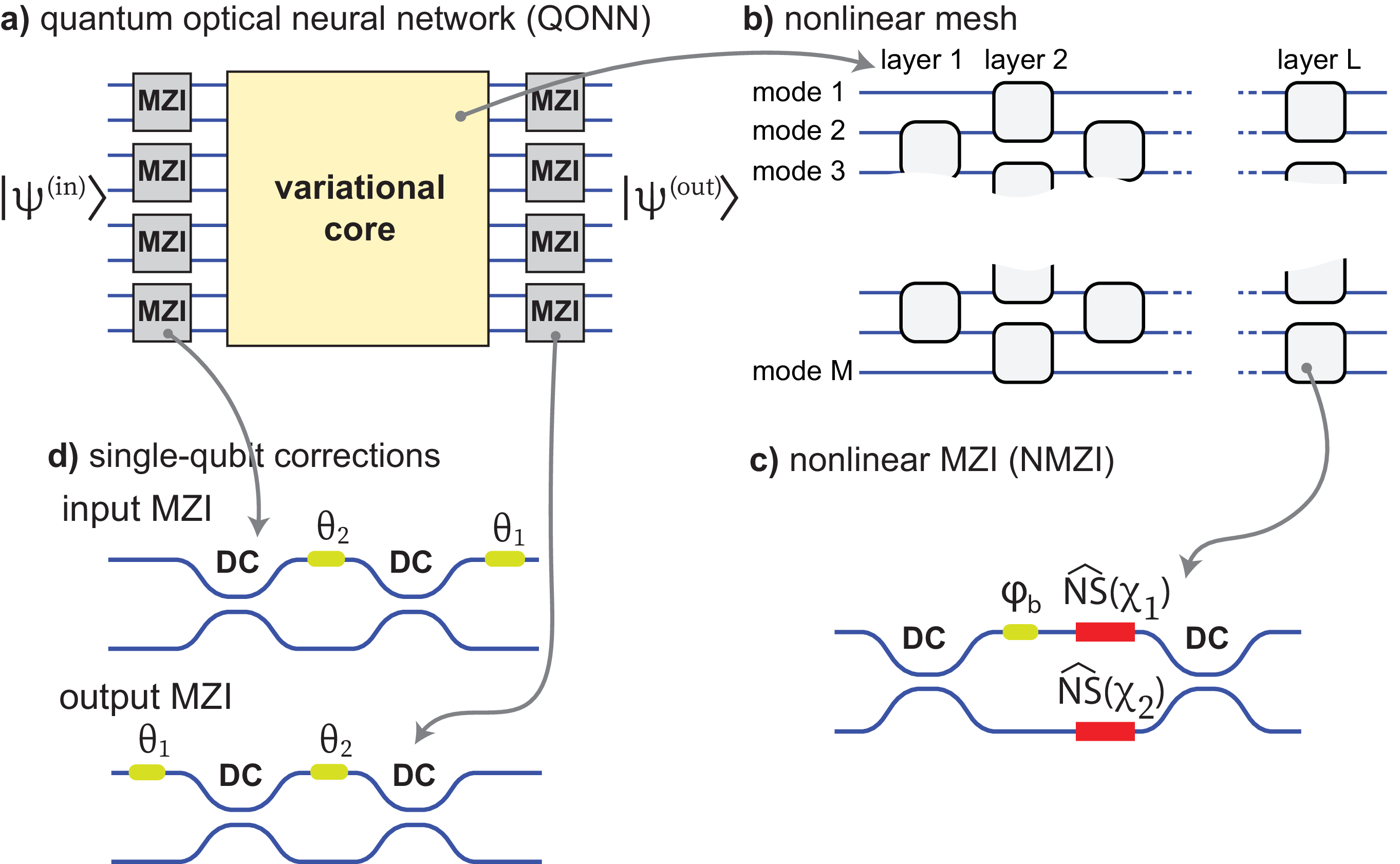}
        \caption{\textbf{The QONN with adjustable nonlinearities}: a) QONN performing transformation of the dual-rail-encoded states. The QONN consists of a programmable nonlinear multimode network followed by programmable single-qubit operations performed; b) The $M$-mode nonlinear network constructed by a $L$-layer mesh of two-mode nonlinear Mach-Zehnder interferometers (NMZI); c) the circuit of a single NMZI consisting of two balanced directional couplers (DCs), static linear phase-shift bias $\varphi_b$ and two nonlinear sign-shift gates (NS-gates) in the arms individually programmed by parameters $\chi_1$ and $\chi_2$; d) the circuit of a Mach-Zehnder interferometer programmed by the phase shifts $\theta_1$ and $\theta_2$. The trained QONN parameters include the set of the NS-gate parameters $\boldsymbol{\chi}$ and output phase-shifts $\boldsymbol{\theta}$.}\label{fig:qonn_architecture}
    \end{figure*}

The proposed QONN is depicted in Fig.~\ref{fig:qonn_architecture}. The main functional block of the QONN is the variational core shown in Fig.~\ref{fig:qonn_architecture}a --- a programmable nonlinear $M$-mode network transforming  quantum states of photons. The QONNs under study operates in the Hilbert space ${\cal{}H}_N^M$ of discrete $N$-photon states occupying $M$-modes, in which the states are superpositions of the form $\sum_{\boldsymbol{t}\in\Omega_N^M}c_{\boldsymbol{t}}|\boldsymbol{t}\rangle$, where $\boldsymbol{t}=(t_1,t_2,\ldots,t_M)$ is the photon-occupation vector designating a particular $N$-photon Fock state $|\boldsymbol{t}\rangle$ ($\sum_{m=1}^M{}t_m=N$) and the corresponding probability amplitude $c_{\boldsymbol{t}}$, $\Omega_N^M$ is the set of all possible $N$-photon $M$-mode Fock states with dimension $\text{dim}{\cal{}H}_N^M=\left(\begin{array}{c}
    M+N-1 \\
     N
\end{array}\right)$. In the following, we will be interested in generation of transformation of the dual-rail-encoded states, so that throughout the work $M=2N$ and the number of photons $N$ will also designate the number of qubits.

Specifically, we study a mesh of two-mode blocks shown in Fig.~\ref{fig:qonn_architecture}b. The two-mode blocks   are nonlinear interferometers shown in Fig.~\ref{fig:qonn_architecture}c. In the rest of the paper we call the two-mode nonlinear block a nonlinear Mach-Zehnder interferometer (NMZI), due to its similarity with the traditional linear optical MZI. The NMZIs are constructed from the  balanced directional couplers (DCs) characterized by the transfer matrices $U_{\text{DC}}=\frac{1}{\sqrt{2}}\left(\begin{array}{cc}
    1 & 1 \\
    1 & -1
\end{array}\right)$, and two programmable nonlinear sign-shift gates (NS-gates) placed in the arms between the DCs.  The programmable NS-gate exploits the Kerr-like nonlinearity described by the following operator: 
    \begin{equation}\label{eqn:nonlinearity}
        \hat{\text{NS}}_m(\chi)=\exp\left(i\frac{\chi}{2}(\hat{a}_m^{\dagger})^2\hat{a}_m^2\right),
    \end{equation}
where $\hat{a}_m$ ($\hat{a}_m^{\dagger}$) is the photon annihilation (creation) operator acting on the mode with index $m$. The parameter $\chi$ has the meaning of nonlinearity strength, which can take values in the range $0\le\chi\le\pi$. Notice that the programmable nonlinearities are a significant departure from previous works on quantum optical circuits, where the nonlinearities were assumed static~\cite{QuantumOpticalNNs,EwaniukQONNs}. Owing to this property, the QONN under study derives its advantages.

In addition to programmable nonlinearities $\chi$, the NMZIs also include linear phase-shift biases $\varphi_b$. The values of $\varphi_b$ control the linear part of the quantum state evolution passing through the NMZIs, i.\,e. single-photon components of the states, as the nonlinear gate \eqref{eqn:nonlinearity} has no effect on such states. We assume that the values of $\varphi_b$ are set prior to training the variational core circuit, in other words, they are static in the course of running the optimization algorithm. However, later, we will show that the choice of the linear phase-shift biases $\varphi_b$ has a profound effect on the performance of the QONNs. While, obviously, including the linear biases $\varphi_b$ into the space of adjustable parameters on par with the nonlinearities $\mathbf{\chi}$ should make the QONN more capable, we leave the study of linear-optics role in QONNs to another work for us. Therefore, the transformation of the variational core consisting of $L$ NMZI layers is described by the following operator:
    \begin{equation}\label{eqn:Ucore}
        \hat{U}_{\text{core}}(\boldsymbol{\chi})=\prod_{l=1}^L\ \prod_{j\in\Omega_l^{\text{NMZI}}}\hat{U}_{\text{NMZI},j}^{(l)}(\chi^{(l)}_{2j-1},\chi^{(l)}_{2j}),
    \end{equation}
where $\hat{U}_{\text{NMZI},j}^{(l)}(\chi_1,\chi_2)$ is the operation of a single NMZI in the $l$-th layer between modes $j$ and $j+1$, $\Omega_l^{\text{NMZI}}$ denotes the ordered sequence of NMZIs in the layer with index $l$. We write the explicit form of the constituent transformation of a single NMZI $\hat{U}_{\text{NMZI},j}^{(l)}(\chi_1,\chi_2)$ in Appendix \ref{sec:nmzi_transformation}.

\begin{figure}[hbtp]
        \centering\includegraphics[width=0.3\textwidth]{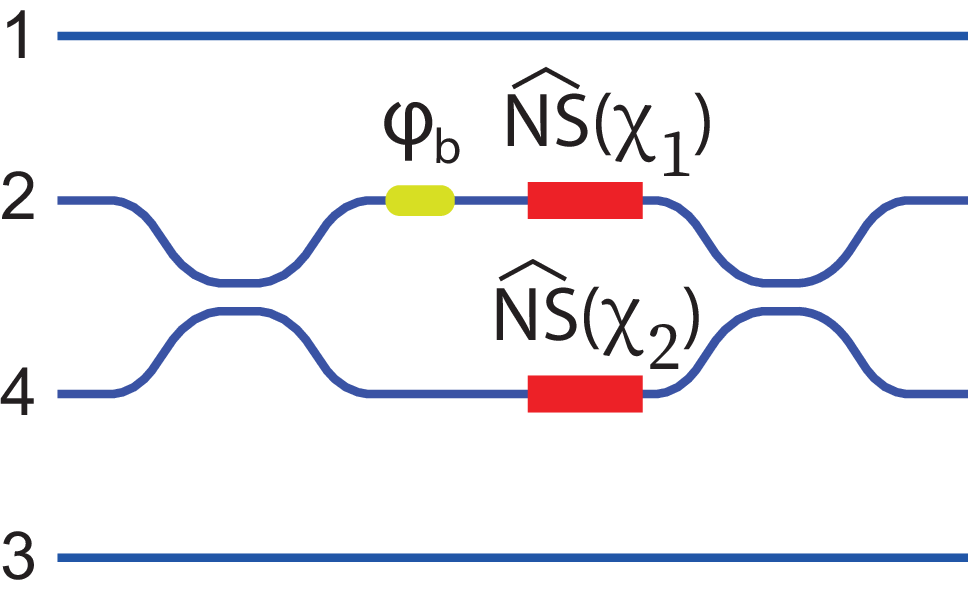}
        \caption{\textbf{Four-mode circuit with a programmable NMZI}. The order of mode indices corresponds to the CZ-gate on dual-rail-encoded qubits at $\chi_1=\chi_2=\pi$ and $\varphi_b=0$.}\label{fig:czgate}
    \end{figure}

Without nonlinearities in the NMZI blocks  the class of possible transformations is limited to the one of multimode \emph{linear} optical networks, which is a very narrow class of operations in the photonic Hilbert space. Generally, NMZI transformations operating in the full Hilbert space result in states that are incompatible with single- and dual-rail qubit encodings which are standard methods of photonic information encoding. 
Also, the NMZI cannot perform single-qubit operations due to non-adjustable $\varphi_b$. For example, Fig.~\ref{fig:czgate} illustrates the four-mode optical circuit utilizing an NMZI. At specific parameter values, $\chi_1=\chi_2=\pi$ and $\varphi_b=0$, this is simply the CZ-gate~\cite{KLM} acting on dual-rail-encoded states at the circuit inputs. However, other choices of $\chi_1$, $\chi_2$, and $\varphi_b$ lead to deviations from the dual-rail qubit encoding. In this work, we show that multiple NMZIs constituent the nonlinear variational core can perform qubit operations.

%

\subsection{Linear optical single-qubit corrections}\label{sec:LO_corrections}

In the examples used to study the capabilities of the proposed approach that follows, one deals with preparation and transformation of dual-rail-encoded qubit states~\cite{Slussarenko}. In the dual-rail encoding, the state of a qubit is a superposition of one photon over two modes, so that the basis qubit states $|0\rangle_L=|1\rangle_1|0\rangle_2$ and $|1\rangle_L=|0\rangle_1|1\rangle_2$, where $|0\rangle_j$ and $|1\rangle_j$ are Fock states of $j^{th}$ mode with $0$ and $1$ photon, respectively.

In this encoding method, arbitrary single-qubit operations are implemented  deterministically by the programmable linear optical MZIs, which are much easier to realize in practice than the nonlinear counterparts --- NMZIs. Taking into account that the phase-shift biases $\varphi_b$ in the NMZIs are non-adjustable and the NS-gates are not susceptible to the single-photon qubit states,  the linear optical MZI layers are added to the input and output of the nonlinar variational core, in order to perform single-qubit corrections.

Therefore, as a whole, our QONN performs the following $M$-mode transformation:
    \begin{equation}
        \hat{U}_{\text{QONN}}(\boldsymbol{\chi},\boldsymbol{\theta})=\hat{U}_{\text{MZI}}^{(out)}(\boldsymbol{\theta}^{(out)})\hat{U}_{\text{core}}(\boldsymbol{\chi})\hat{U}_{\text{MZI}}^{(in)}(\boldsymbol{\theta}^{(in)}),
    \end{equation}
where $\hat{U}_{\text{MZI}}(\boldsymbol{\theta})=\prod_{q=1}^N\hat{U}_{\text{MZI}}^{(in,q)}(\theta_1^{(q)},\theta_2^{(q)})$ with $\hat{U}_{\text{MZI}}^{(q)}$ being the MZI operator acting on a pair of modes encoding a qubit with index $q$. The photonic circuit of MZIs implementing single-qubit corrections is shown in Fig.~\ref{fig:qonn_architecture}d. It is specified by the transfer matrix connecting input and output field operators:
\begin{equation}
        \begin{split}
        U_{\text{MZI}}(\theta_1,\theta_2)=
        U_{\text{DC}}
        \left(
        \begin{array}{cc}
           e^{i\theta_2}  & 0 \\
            0 & 1
        \end{array}
        \right)U_{\text{DC}}
        \left(
        \begin{array}{cc}
           e^{i\theta_1}  & 0 \\
            0 & 1
        \end{array}
        \right) =\\
        = e^{i\theta_{2}/2}
        \left(
        \begin{array}{cc}
           e^{i \theta_1} \cos(\theta_2/2) &  i\sin(\theta_2/2) \\
            i e^{i \theta_1} \sin(\theta_2/2) & \cos(\theta_2/2)
        \end{array}
        \right)
        \end{split}
    \end{equation}
where $\theta_{1,2}^{(q)}$ are tunable phase-shifts that program the MZI corresponding to qubit $q$.

The single-qubit corrections have a substantial impact on reducing the depth of the nonlinear variational core, as they alleviate the need for single-qubit operations, which are incompatible with the NMZIs. However, it is important to emphasize that while programmable linear optics are included in our QONN, their role remains secondary.

\subsection{QONN trained by multimode linear optics}

    \begin{figure*}[htp]
        \centering\includegraphics[width=0.8\textwidth]{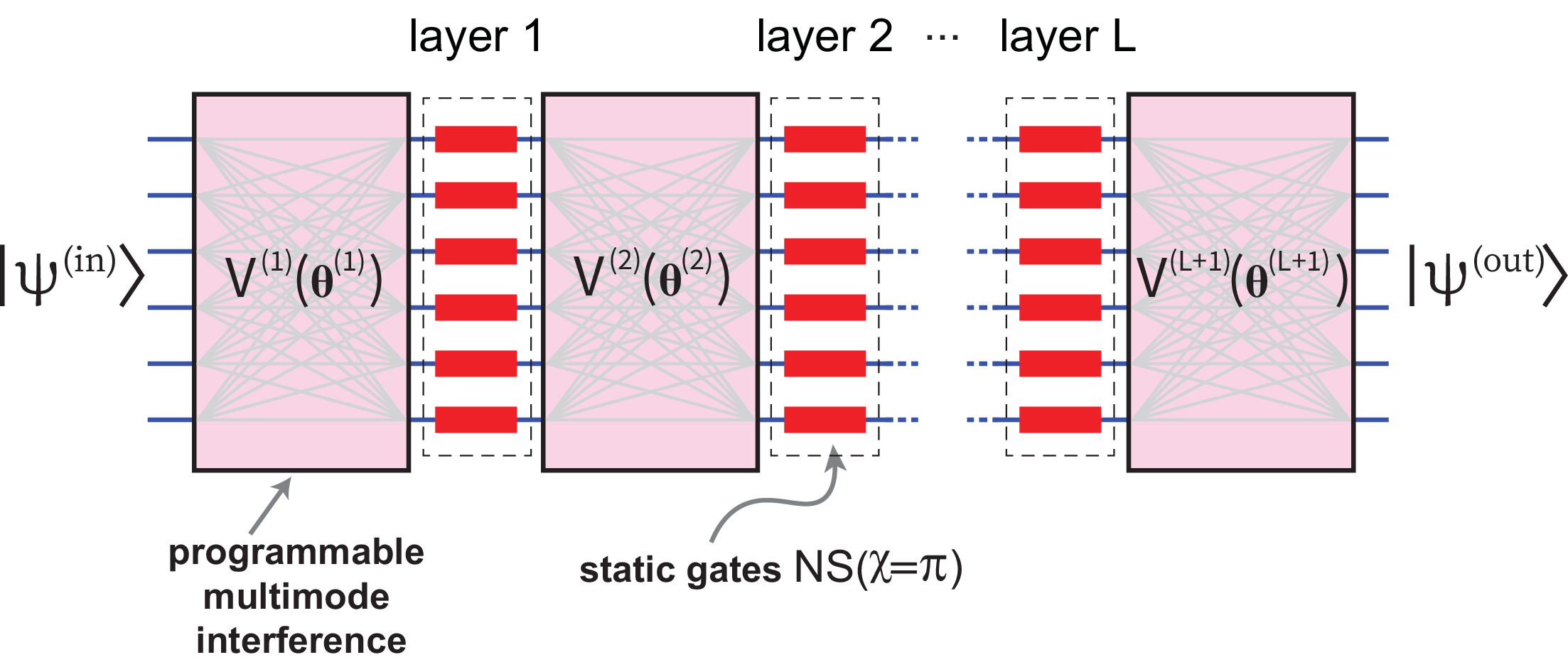}
        \caption{\textbf{The QONN architecture trained by programming linear optics}. The $L$-layer QONN consists of a sequence of alternating programmable multimode interferometers, designated by the corresponding transfer matrices $V^{(l)}$  ($1\le{}l\le{}L+1$), and layers of static NS-gates with maximal nonlinearity strength $\chi=\pi$. The quantum network is trained solely by programming the parameters $\boldsymbol{\theta}^{(l)}$.}\label{fig:lo_architecture}
    \end{figure*}

To evaluate the performance of the proposed QONNs with programmable nonlinearities, we compare them against QONNs trained solely using linear optics~\cite{QuantumOpticalNNs, EwaniukQONNs}. The architecture of such QONNs is shown in Fig.~\ref{fig:lo_architecture}.

The QONN trained with linear optics consists of an interferometer with $M$ input and output modes, followed by $M$ static NS-gates with fixed nonlinearities $\chi=\pi$. The linear optical transformation is represented by transfer matrices $V^{(l)}(\boldsymbol{\theta}^{(l)})$, which are programmed through a set of phase shifts $\boldsymbol{\theta}^{(l)}$ ($l=1,\dots,L+1$). For this analysis, we assume the universality of the interferometers, meaning they can be programmed to implement any transfer matrix by appropriately setting the parameters $\boldsymbol{\theta}^{(l)}$. Several methods exist for designing such universal interferometers~\cite{ReckDesign, ClementsDesign, RobustSaygin, FldzhyanDesign}. In our case, we used the Clements design~\cite{ClementsDesign}, which requires $M^2$ phase-shift parameters for each $M$-mode interferometer.

\subsection{Relevant parameters of QONNs}

Currently, integrated photonics technologies represent a promising avenue for the development of quantum optical circuits~\cite{IntegratedPICreview,QDforQIP}. Several key components of these circuits have already been demonstrated on-chip, including photon sources~\cite{LodahlSource,SourceIntegrated}, photodetectors~\cite{IntegratedDetectors}, programmable linear optics~\cite{Taballione2023modeuniversal}, and single-photon-sensitive nonlinearities~\cite{SinglePhoton_NL_nature,Lodahl_arxiv_NS}. However, only a subset of these elements have been integrated into a single platform, and much work remains to fully integrate all the necessary components for large-scale quantum optical circuits.

Our work advances the scalability of programmable quantum optical circuits, which typically consist of two main components: multimode linear optics, commonly implemented using MZI meshes, and single-photon-sensitive nonlinearities. In practical implementations, each type of the components introduces different degrees of losses, errors, and noise, and each poses different fabrication challenges. While it is generally desirable to minimize the number of nonlinear layers, analyzing the impact of the nonidealities introduced by these components in a QONN is complex and requires specific details about the underlying platform. 

Linear MZI meshes are relatively simple to fabricate, as they consist of standard elements such as static directional couplers and programmable phase-shifters~\cite{ClementsDesign}. However, constructing even moderately sized universal programmable linear-optical circuits remains challenging due to their large size and relatively slow modulation speeds~\cite{Taballione2023modeuniversal,CryoPS}. Recent proposals suggest overcoming these limitations by replacing traditional phase-shifting elements with quantum dots as reconfigurable phase shifters~\cite{QDphaseshifts}. This approach aims to eliminate the need for bulky linear phase shifters, albeit at the cost of more complex fabrication and control.

Single-photon-sensitive nonlinearities can be realized using single emitters, such as quantum dots, embedded in photonic waveguides or cavities. The strong confinement of light in these structures enables near-unity light-matter coupling efficiencies. Recently, a nonlinear phase shift was demonstrated using a single quantum dot within a planar nanophotonic waveguide~\cite{Lodahl_arxiv_NS}. It is worth noting that making nonlinear elements programmable is no more difficult than realizing static nonlinear elements. Since the variable parameter of a nonlinearity is determined by the coupling strength, it can be adjusted post-fabrication, for example, by detuning the emitter’s transition frequency or the resonant frequency of the cavity.

To highlight the advantages of our QONN architecture, we focus on two key characteristics of quantum optical circuits: (1) the circuit depth $D$ and (2) the number of adjustable parameters $P$ required to achieve high-quality performance. Specifically, we examine the linear-optical circuit depth $D_{\text{LO}}$, defined as the number of linear phase-shift layers in the variational core, and the nonlinear circuit depth $D_{\text{NL}}$, which corresponds to the number of nonlinear layers. It is important to note that the proposed nonlinearity-adjustable QONNs have $D_{\text{LO}}=0$, as they do not have programmable linear phase shifts in their variational core. In contrast, traditional linear-optics-adjustable QONNs have both $D_{\text{LO}}$ and $D_{\text{NL}}$ greater than zero, and these two parameters are interrelated. For instance, assuming a compact version of the universal MZI meshes with the minimum number of phase-shift layers~\cite{Bell}, the relationship between $D_{\text{LO}}$ and $D_{\text{NL}}$ and is given by:
    \begin{equation}\label{eqn:DLO_DNL}
        D_{\text{LO}}=(M+1)(D_{\text{NL}}+1).
    \end{equation}
In the following, we use $D_{\text{NL}}$ to quantify the circuit depths for both QONN architectures and additionally compute $D_{\text{LO}}$ for the linear-optics-adjustable QONNs using the above equation.

The number of parameters $P$ in a quantum neural network is related to the computational resources required for training. Increasing the number of parameters can make training more challenging due to the curse of dimensionality, which is associated with the barren plateau phenomenon~\cite{BarrenPlateaus,BarrenPlateausReview}. However, the difficulty of training is more nuanced, and it largely depends on the architecture of the quantum circuit. A detailed analysis of the trainability of QONNs will be conducted in future work. The scaling of the number of adjustable parameters $P_{\text{NL}}/P_{\text{LO}}$ with circuit depth $D_{\text{NL}}$ differs significantly between the proposed nonlinearity-adjustable QONNs and traditional linear-optics-adjustable QONNs. Specifically, adding one nonlinear layer to the former increases the number of adjustable parameters by MM, while in the latter, the increase is proportional to $M(M-1)$. As a result, the total number of adjustable parameters for the nonlinearity-adjustable QONN is given by $P_{\text{NL}}= D_{\text{NL}} \times (M-2) + 2\lfloor D_{\text{NL}}/2 \rfloor $ excluding the phase shifts used for single-qubit corrections. For the linear-optics-adjustable QONN, the total number of parameters is $P_{\text{L}}=(D_{\text{NL}}+1) \times M(M-1)$.

\subsection{Relation with gate-based circuits}

With deterministic nonlinearities it is possible to overcome the limitations of linear optics. In particular, deterministic two-qubit gates can be implemented using NS-gates (see Fig.~\ref{fig:czgate}), hence, one can prepare arbitrary qubit states through constructing proper gate circuit~\cite{Kitaev}. However, QONNs differ from gate-based systems in that they utilize the entire Hilbert space $\mathcal{H}_N^M$ spanned by $M$ modes and $N$ photons, and $\dim \mathcal{H}_N^M$ scales more rapidly than the dimension of the corresponding dual-rail-qubit subspace. We also note that in the considered QONN architecture, the linear optical interference within the NMZI mesh does not precisely replicate the interference patterns found in quantum optical gate circuits. Therefore, during the training process of our QONNs, we do not expect to converge to gate-based circuits that perform the same tasks.

\section{Results}\label{sec:results}

To assess  our QONN and compare its performance with the reported QONNs, below we consider some problems relevant for quantum information.

\subsection{Generation of entangled qubit states}\label{sec:generation}

The generation of entangled quantum states is a crucial problem in the field of quantum technologies and quantum computing in particular~\cite{shor1995scheme, MBQC, ekert1991quantum}. Since entangled states are essential for quantum technologies, their preparation is a significant challenge within this broad area. However, generating entangled states using only linear optics is demanding, as it necessitates the use of many sophisticated physical components to address the non-deterministic nature of the state generation process~\cite{bartolucci2021creation}.

We challenge our QONN with the task of preparation of $N$-qubit entangled states. Firstly, the QONN aims to prepare the following state:
        \begin{align}\label{eqn:ghz_parametrized}
            |\text{G}_{N}(\alpha)\rangle=\cos\alpha|0\rangle_L^{\otimes{}{N}}+\sin\alpha|1\rangle_L^{\otimes{}{N}}\nonumber\\
            =\cos\alpha|10\rangle^{\otimes{}{N}}+\sin\alpha|01\rangle^{\otimes{}{N}},
        \end{align}
starting from the initial separable state  $|\Psi^{(in)}\rangle=|10\rangle^{\otimes{}N}$. In \eqref{eqn:ghz_parametrized}, the angle parameter $0\le\alpha\le\pi/4$ specifies the pre-defined entanglement degree of the state. Notice that the entanglement degree is a monotonic function of $\alpha$ with  $\alpha=0$ and $\alpha=\pi/4$ corresponding to the separable and maximally entangled $N$-qubit GHZ states, respectively~\cite{Programmable2qubits}. This fact allows us to analyze the effect of the entanglement degree on the QONN's performance simply by varying $\alpha$ in \eqref{eqn:ghz_parametrized}.

To quantify the QONN performance, we used the standard fidelity cost function:
    \begin{equation}\label{eqn:fidelity}
        F_{\alpha}(\boldsymbol{\chi},\boldsymbol{\theta})=|\langle\Psi^{(out)}(\boldsymbol{\chi},\boldsymbol{\theta})|G_N(\alpha)\rangle|^2,
    \end{equation}
where $|\Psi^{(out)}(\boldsymbol{\chi},\boldsymbol{\theta})\rangle$ is the actual quantum state prepared at given nonlinearities  $\boldsymbol{\chi}$ and single-qubit correction parameters $\boldsymbol{\theta}$. Due to the stringent quality requirements for quantum states to be used in quantum devices, the acceptable fidelity should be as close as possible to the maximum value, $F_{\alpha}(\boldsymbol{\chi}, \boldsymbol{\theta}) = 1$. Note that the ideal case also corresponds to the deterministic generation of the target state, so both state quality and generation probability are captured by the cost function~\eqref{eqn:fidelity}.

For all the problems considered in this work, the networks were trained using a numerical optimization procedure detailed in Appendix~\ref{sec:optimization}. During training, the optimization algorithm searched for a global minimum of a cost function (CF) that quantifies the error in the task under consideration. Specifically, for the state generation problem, the cost function is the infidelity: $CF = 1-F_{\alpha}(\boldsymbol{\chi}, \boldsymbol{\theta})$.  The same optimization procedure was used for QONNs programmed with linear optics and static nonlinearities (see Fig.~\ref{fig:lo_architecture}), allowing us to compare these two approaches to constructing QONNs. We analyzed the effect of depth $D$ on the QONN's capabilities by training the QONN across a range of $D$. Typically, the likelihood of CF converging to a good minimum in a single optimization run is quite low due to the algorithm often falling into a local minimum, which necessitated multiple optimization runs for each target state $|G_N(\alpha)\rangle$ and depth value $D_{\text{NL}}$.

    \begin{figure*}[htp]
        \centering\includegraphics[width=0.85\textwidth]{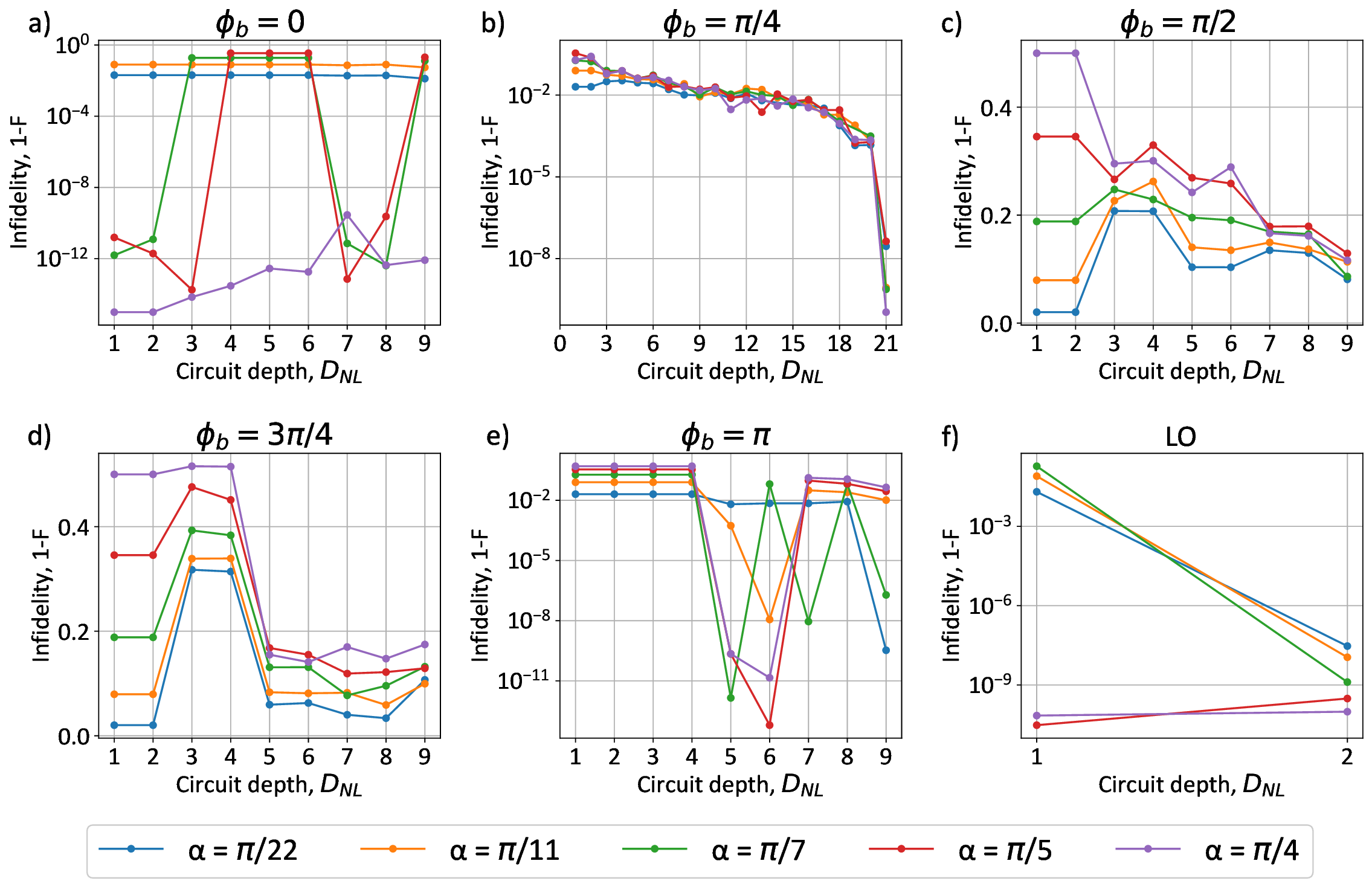}
        \caption{\textbf{Preparation of the $3$-qubit states $|\text{G}_{N}(\alpha)\rangle$ ($N=3$) with various entanglement degree by different QONNs}: state infidelity $1-F$ for the QONN with programmable nonlinearities as a function of circuit depth  $D_{\text{NL}}$ for  various values of $\alpha$ and different biases $\varphi_b=0$ (a), $\pi/4$ (b), $\pi/2$ (c),  $3\pi/4$ (d) and $\pi$ (e). The same dependence for QONN with programmable linear optics is plotted in (f).}\label{fig:ghz3_gen}
    \end{figure*}
    \begin{figure*}[htp]
        \centering\includegraphics[width=0.85\textwidth]{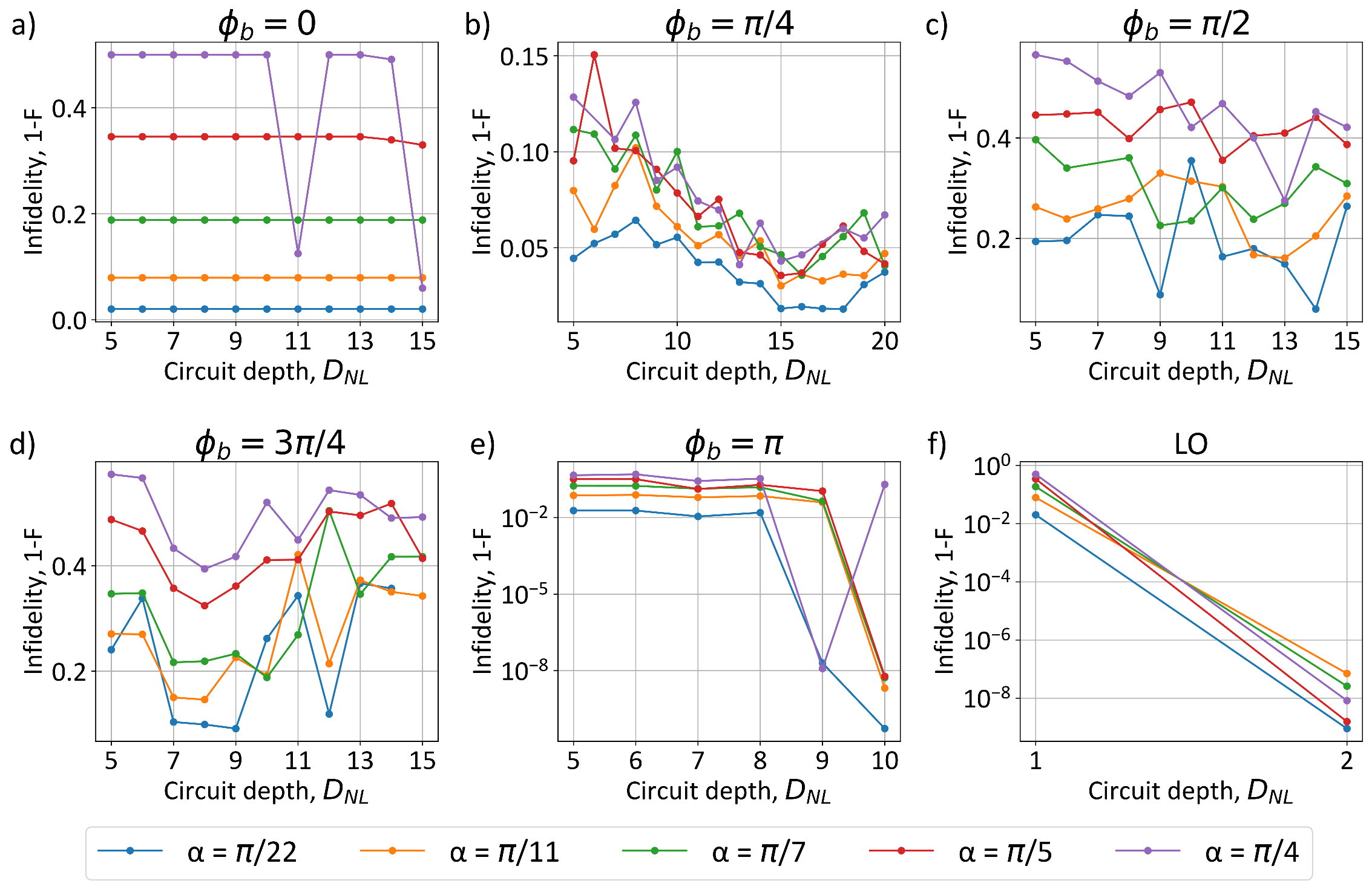}
        \caption{\textbf{Preparation of the $4$-qubit states $|\text{G}_{N}(\alpha)\rangle$ ($N=4$) with various entanglement degree by different QONNs}: state infidelity $1-F$ for the QONN with programmable nonlinearities as a function of circuit depth  $D_{\text{NL}}$ for  various values of $\alpha$ and different biases $\varphi_b=0$ (a), $\pi/4$ (b), $\pi/2$ (c),  $3\pi/4$ (d) and $\pi$ (e). The same dependence for QONN with programmable linear optics is plotted in (f).} \label{fig:ghz4_gen}
    \end{figure*}

Fig.~\ref{fig:ghz3_gen} and Fig.~\ref{fig:ghz4_gen} show the obtained dependencies of infidelity on circuit depth for various values of the state parameter $\alpha$ for $N=3$ and $N=4$ qubits, respectively. In addition to varying $\alpha$, we also examined different phase shifts $\varphi_b$ that control the linear optical part of the QONN transformation. We selected linear optical configurations with $\varphi_b$ values of 0, $\pi/4$, $\pi/2$, $3\pi/4$, and $\pi$ (the corresponding dependencies are drawn in plots a)-e) of the figures). Our QONN is not capable of preparing the states for all presented values of $\varphi_b$ --- only two of these values, specifically $\varphi_b = \pi/4$ and $\varphi_b = \pi$, work. We notice that the dependencies are radically different for these two cases. At $\varphi_b = \pi/4$, the infidelity almost monotonically decreases with increasing circuit depth for all target states, so one predefined depth value $D_{\text{NL}}$ is universal for all the states. In contrast, at $\varphi_b = \pi$ infidelity for the $3$-qubit states is non-monotonic and also each target state requires its own optimal value of $D_{\text{NL}}$.  At the same time, for the $4$-qubit states infidelity drops monotonically and starting from a  value of  $D_{\text{NL}}=10$ all the states can be prepared with high quality.


%
    \begin{figure*}[htbp]
        \centering\includegraphics[width=0.9\textwidth]{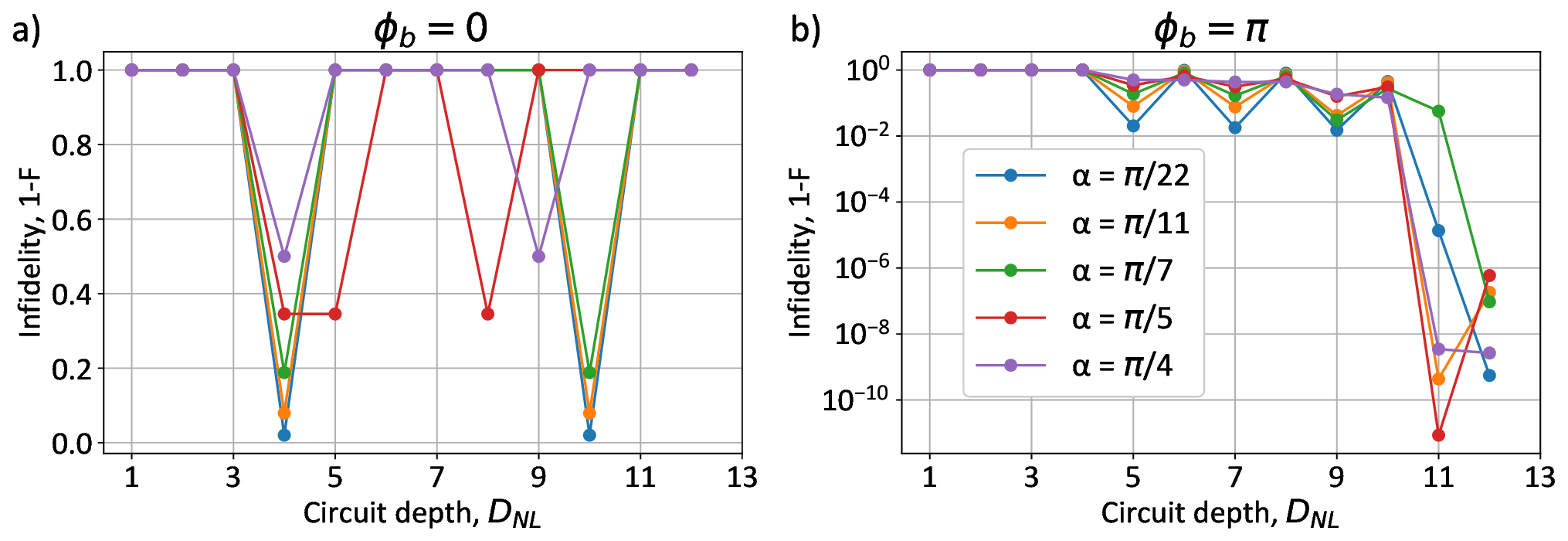}
        \caption{\textbf{Preparation of the $3$-qubit states $|\text{G}_{N}(\alpha)\rangle$ ($N=3$) with various entanglement degree by the linear-optically programmed QONN without single-qubit corrections}: the state infidelity $1-F$ as a function of circuit depth $D_{\text{NL}}$ for entangled states preparation with various parameter $\alpha$ at different  phase-shift biases $\varphi_b = 0$ (a) and $\pi$ (b).}\label{fig:ghz3_gen_noMZI}
    \end{figure*}

The infidelities for the QONN programmed by linear optics are also plotted in Fig.~\ref{fig:ghz3_gen}f and Fig.~\ref{fig:ghz4_gen}f in order to compare their performance with our QONNs. As one can see, the QONN programmed by linear optics is capable of preparing the states as well. Moreover, it requires a lower number of (static) nonlinear layers $D_{\text{NL}}$ than in our QONNs. Importantly, however, that the number of phase-shift layers $D_{\text{LO}}$ and the total number of adjustable phase-shifts $P_{\text{LO}}$ required to prepare the states with high fidelity scales is much larger, than the number of nonlinear phase-shift layers $D_{\text{NL}}$ and the total number of adjustable nonlinear phase-shifts $P_{\text{NL}}$ in our QONN.

In particular, for maximally entangled states of $N=3,4$ qubits the parameter counts reads $P_{\text{NL}} = 4, 62$ and $P_{\text{LO}}= 60, 168$ for our QONN and the counterpart programmed by linear optics, respectively. Table~\ref{tab:ghz_gen_table} summarizes the obtained QONN parameters. The much smaller nonlinear depth $D_{\text{NL}}$ of our QONN than the linear depth $D_{\text{LO}}$ of the counterpart QONN suggests that it can be implemented on a  compact quantum circuit.

\begin{table}
\centering
\caption{Nonlinear $D_{\text{NL}}$ and linear circuit depth  $D_{\text{LO}}$ and parameter count $P$ for QONNs peforming maximally entangled state generation (see Fig.~\ref{fig:ghz3_gen},~\ref{fig:ghz4_gen}). NL and LO are the QONNs with adjustable nonlinearities and the counterparts trained by linear optics, respectively.}
\begin{tblr}{
  cell{1}{1} = {r=2}{},
  vlines,
}
\hline
 &  $D_{\text{NL}}$ & $D_{\text{LO}}$  & $P$ \\
\hline
 & $3$-/$4$-qubit &  $3$-/$4$-qubit  & $3$-/$4$-qubit \\
\hline
NL ($\varphi_b=0$)  & 1/-- &  0 & 4/--  \\
NL ($\varphi_b=\pi$) & 5/9 &  0 & 24/62  \\
LO  & 1/2 &  14/27 & 60/168  \\
\hline
\end{tblr}\label{tab:ghz_gen_table}
\end{table}

    \begin{figure}[htbp]
        \centering\includegraphics[width=0.48\textwidth]{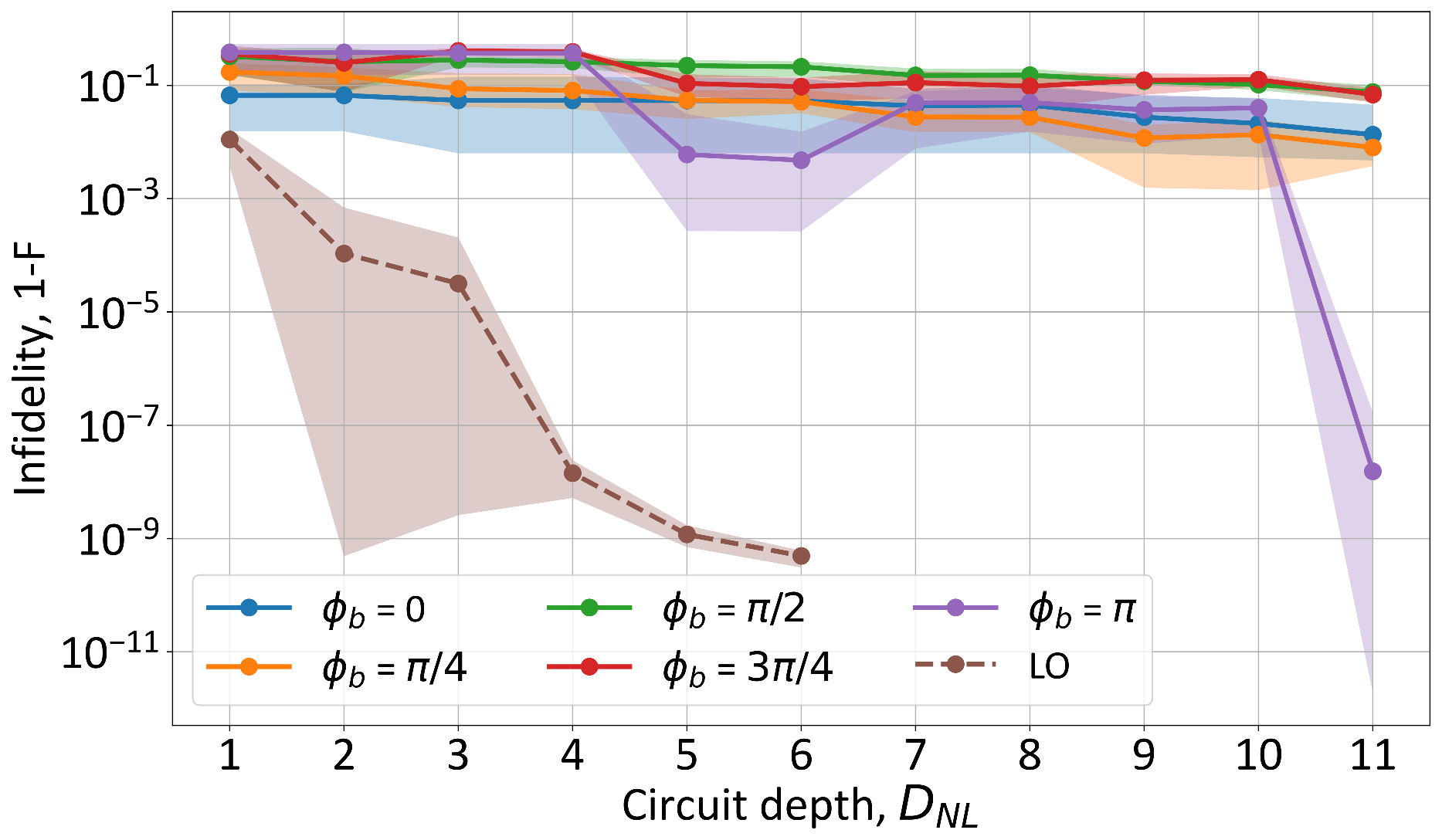}
        \caption{\textbf{Preparation of randomly sampled $3$-qubit states}: the state infidelity $1-F$ as a function of circuit depth  $D_{\text{NL}}$ at different values of linear phase-shift biases $\varphi_b$ for the proposed QONN (solid curves) and for the QONN programmed by linear optics (dashed curve). The curves have been obtained for $30$ randomly generated states.} \label{fig:random3_gen}
    \end{figure}

Additionally, we studied separately the role of single-qubit corrections on QONN performance. For this, we fixed the MZIs' parameters $\boldsymbol{\theta}$ in such a way that they perform identity operations, and run the optimization algorithm now exploring only the variational core's nonlinearities $\boldsymbol{\chi}$. Fig.~\ref{fig:ghz3_gen_noMZI} shows the corresponding infidelities for the two biases $\varphi_b = 0$ and $\varphi_b = \pi$ with the best results in the previous calculations for a network with programmable MZI.
Although the QONN is capable of getting the job done without resorting to the single-qubit corrections, it costs several additional layers compared with the QONN having single-qubit corrections. Therefore, taking into account that the linear MZIs are more easier to implement than the nonlinear elements, the utilization of the single-qubit corrections are justified.

The quantum states \eqref{eqn:ghz_parametrized} described above represent a very limited class of all possible $N$-qubit states. Thus, we consider more generic $N$-qubit states with all complex amplitudes being nonzero, i.\,e. $|\Psi_{\text{rnd}}\rangle=\sum_{\boldsymbol{q}}c_{\boldsymbol{q}}|\boldsymbol{q}\rangle$, where $\boldsymbol{q}$ designates the photon occupation vector that run through all $2^n$ states $|\boldsymbol{q}\rangle$ of $N$ dual-rail-encoded qubits and $c_{\boldsymbol{q}}$ denotes the complex amplitude for the state $|\boldsymbol{q}\rangle$ ($\sum_{\boldsymbol{q}}|c_{\boldsymbol{q}}|^2=1$). In particular, we derive the target states by randomly sampling the complex-valued unit vector of $c_{\boldsymbol{q}}$ from a Haar-uniform distribution \cite{RandomPointsnSphere} and analyze the QONN performance as done above. Due to computational complexity of the task we were only able to simulate QONN with $N=3$ qubits. The infidelity dependencies for both our QONN and the QONN with programmable linear optics is shown in Fig.~\ref{fig:random3_gen}.

It is apparent from the figure that the uniformly-sampled states can be prepared by the QONN, however, now it should be deeper than for states \eqref{eqn:ghz_parametrized}. This is due to the fact that the Haar-random target states cover the entire $N$-qubit Hilbert space, in contrast to the states \eqref{eqn:ghz_parametrized} occupying only a subspace of dimension two. This results in deeper circuits for the randomly-sampled states relative to the parametrized states \eqref{eqn:ghz_parametrized} from a narrower class.

It is instructive to compare the number of adjustable parameters required by the QONNs relative to the values set by circuits capable of preparing arbitrary qubit states. One can use the following obvious reasoning: a generic $N$-qubit state can be parameterized by $2^{N+1}-1$ real parameters (if common phase is not included), suggesting that a parameterized quantum circuit designed to prepare arbitrary states should include at least that much programmable parameters. In particular, for $N=3$ the universal circuit should contain  $P\ge{}15$ parameters, respectively. At the same time, using the results depicted in Fig.~\ref{fig:random3_gen}, one obtains $P_{\text{NL}}=54$ and $P_{\text{LO}} = 150$ for our and the linear-optically programmed QONN, respectively. This suggests that both types of parameterized quantum optical circuits are far suboptimal from the gate-based circuits. However, using programmable nonlinearities rather than programmable linear optics turns out to be beneficial for the state generation task.

Table~\ref{tab:rand_state_gen_table} summarizes the QONN circuit parameters obtained. As before, the depth of our QONN with adjustable nonlinearities turned out to be smaller, than the depth of the QONN programmed by linear optics.

\begin{table}
\centering
\caption{Nonlinear $D_{\text{NL}}$ and linear circuit depth  $D_{\text{LO}}$ and parameter count $P$ of QONNs peforming $3$-qubit state generation (see Fig.~\ref{fig:random3_gen}). NL and LO are the QONNs with adjustable nonlinearities and the counterparts trained by linear optics, respectively.}
\begin{tblr}{
  vlines,
}
\hline
 &  $D_{\text{NL}}$ &  $D_{\text{LO}}$ &  $P$  \\
 \hline
NL ($\varphi_b=\pi$) & 11 &  0 & 54  \\
LO  & 4 &  35 & 150  \\
\hline
\end{tblr}\label{tab:rand_state_gen_table}
\end{table}

%
%

\subsection{Bell-state discrimination}\label{sec:discrimination}

Next, we challenge our QONN by training it to implement the deterministic Bell-state discrimination. An ideal Bell-state discriminator is capable to deterministically and unambiguously analyze the following four quantum states:
%
        \begin{eqnarray}\label{eqn:bell_states}
            |\Phi^{(\pm)}\rangle=\frac{|0\rangle_{L1}|0\rangle_{L2}+|1\rangle_{L1}|1\rangle_{L2}}{\sqrt{2}}=\frac{|1010\rangle\pm|0101\rangle}{\sqrt{2}},\nonumber\\
            |\Psi^{(\pm)}\rangle=\frac{|0\rangle_{L1}|1\rangle_{L2}+|1\rangle_{L1}|0\rangle_{L2}}{\sqrt{2}}=\frac{|1001\rangle\pm|0110\rangle}{\sqrt{2}}, 
        \end{eqnarray}
%
where the right-hand side are the Bell-states encoded by $N=2$ photons and $M=4$ modes. The discriminator can transform each of the Bell-states $|\Psi^{(in)}_j\rangle\in\{|\Phi^{(+)}\rangle,|\Phi^{(-)}\rangle,|\Psi^{(+)}\rangle,|\Psi^{(-)}\rangle\}$ into a multimode Fock state $|\Psi^{(out)}_j\rangle\in\{|1010\rangle,|1001\rangle,|0110\rangle,|0101\rangle\}$. In turn, measuring these Fock states by photodetection unambiguously tells about a corresponding input state.

In addition to the standard four Bell states~\eqref{eqn:bell_states}, one can consider the following photonic states:
    \begin{equation}\label{eqn:extrastates_pm}
        |\Theta^{(\pm)}\rangle=\frac{|0011\rangle\pm|1100\rangle}{\sqrt{2}}, 
    \end{equation}
which are equally useful. Although states \eqref{eqn:extrastates_pm} do not respect the dual-rail encoding, they can still serve as a maximally entangled resource that fueled a quantum algorithm, provided cheap linear transformation is performed. Namely, to convert these states into any of the Bell states, one can apply simple two-mode operations, implemented deterministically by linear optics~\cite{bartolucci2021creation}. Therefore, in the following we consider the states \eqref{eqn:extrastates_pm} on par with the Bell states \eqref{eqn:bell_states}. We demand that the discriminator convert the input states $|\Theta^{(+)}\rangle$ and $|\Theta^{(-)}\rangle$ into $|\Psi^{(out)}_5\rangle=|0011\rangle$, $|\Psi^{(out)}_6\rangle=|1100\rangle$, respectively.

It is impossible to obtain a deterministic discrimination of Bell states using  linear optics only. In particular, without using auxiliary photons, the maximal discrimination probability cannot exceed $50$\%~\cite{LO_BELL_DISCRIMINATOR}. The probability can be improved at the cost of extra photons, but it is still far from deterministic~\cite{Busted_discriminator}. Therefore, exploitation of deterministic nonlinearities come in handy to accomplish this task.

To train our neural network as an $S$-state discriminator, we used the following cost function: 
    \begin{equation}\label{eqn:cf_discriminator}
        CF_{disc}=\frac{1}{S}\sum_{j=1}^S|\langle{}\Psi^{(out)}_j|\tilde{\Psi}_j(\boldsymbol{\chi})\rangle|^2,
    \end{equation}
that takes its maximum value of $1$ if all the output states $|\tilde{\Psi}_j(\boldsymbol{\chi})\rangle$ generated by the QONN from input states $|\Psi^{(in)}_j\rangle$ perfectly equal to the corresponding target states $|\Psi^{(out)}_j\rangle$. We trained separately a) the four-state discriminator (states $|\Psi^{(\pm)}\rangle$ and $|\Phi^{(\pm)}\rangle$)  and b) the six-state discriminator (states $|\Psi^{(\pm)}\rangle$, $|\Phi^{(\pm)}\rangle$ and $|\Theta^{(\pm)}\rangle$).

%
%

%
    \begin{figure}[htp]
        \centering\includegraphics[width=0.45\textwidth]{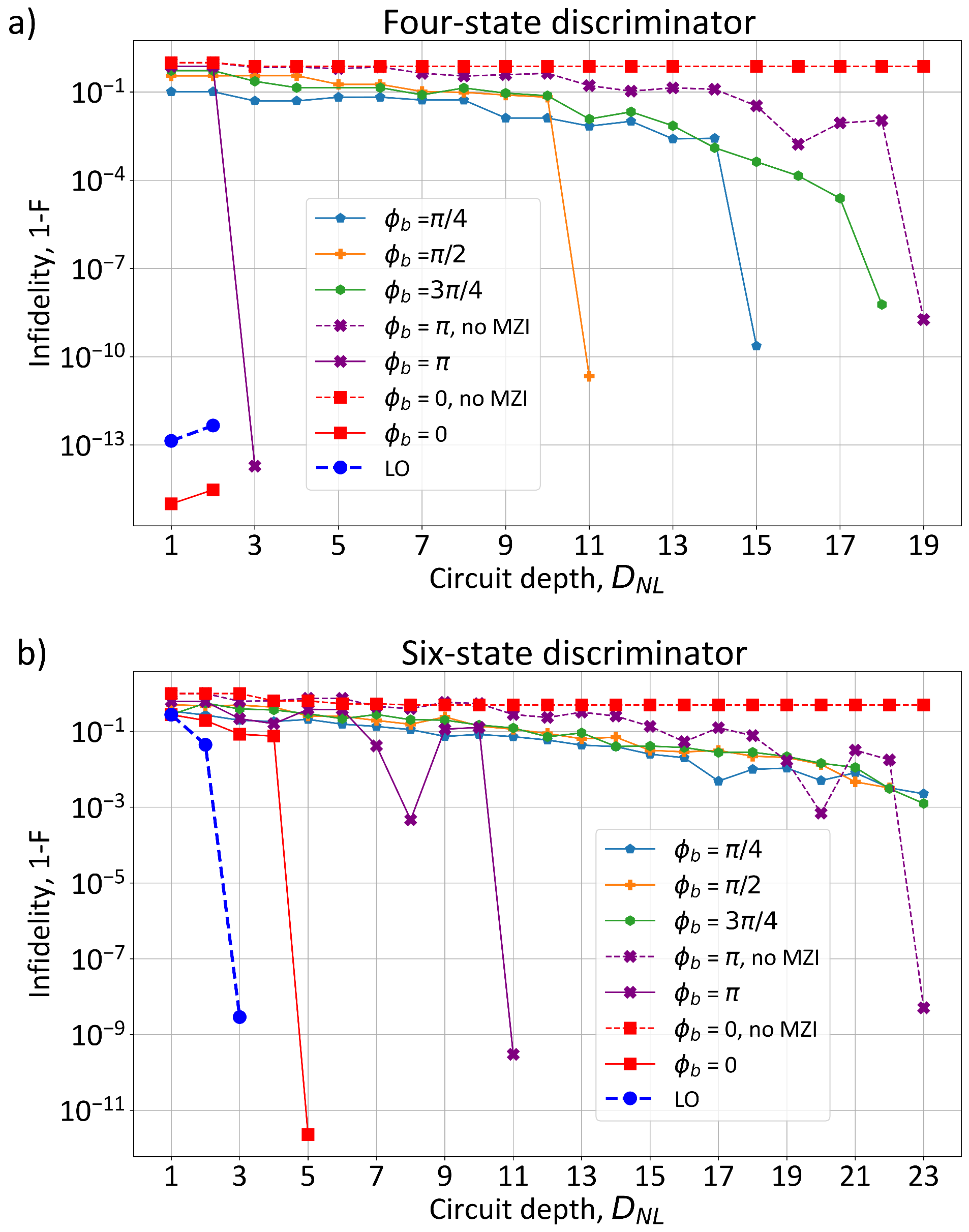}
        \caption{\textbf{Bell-state discrimination performed by the QONNs}: state infidelity $1-F_{disc}$ as a function of circuit depth  $D_{\text{NL}}$ for four- (a) and six-state discrimination (b) at different static phase-shift biases $\varphi_b=0$ of the QONN programmed by nonlinearities (solid curves). The dependencies corresponding to the QONN programmed by linear optics is plotted by the dashed curves.   Each point is the best result of the optimization algorithm run $50$ times.}\label{fig:discriminators_infidelity}
    \end{figure}
    \begin{figure}[htp]
        \centering\includegraphics[width=0.48\textwidth]{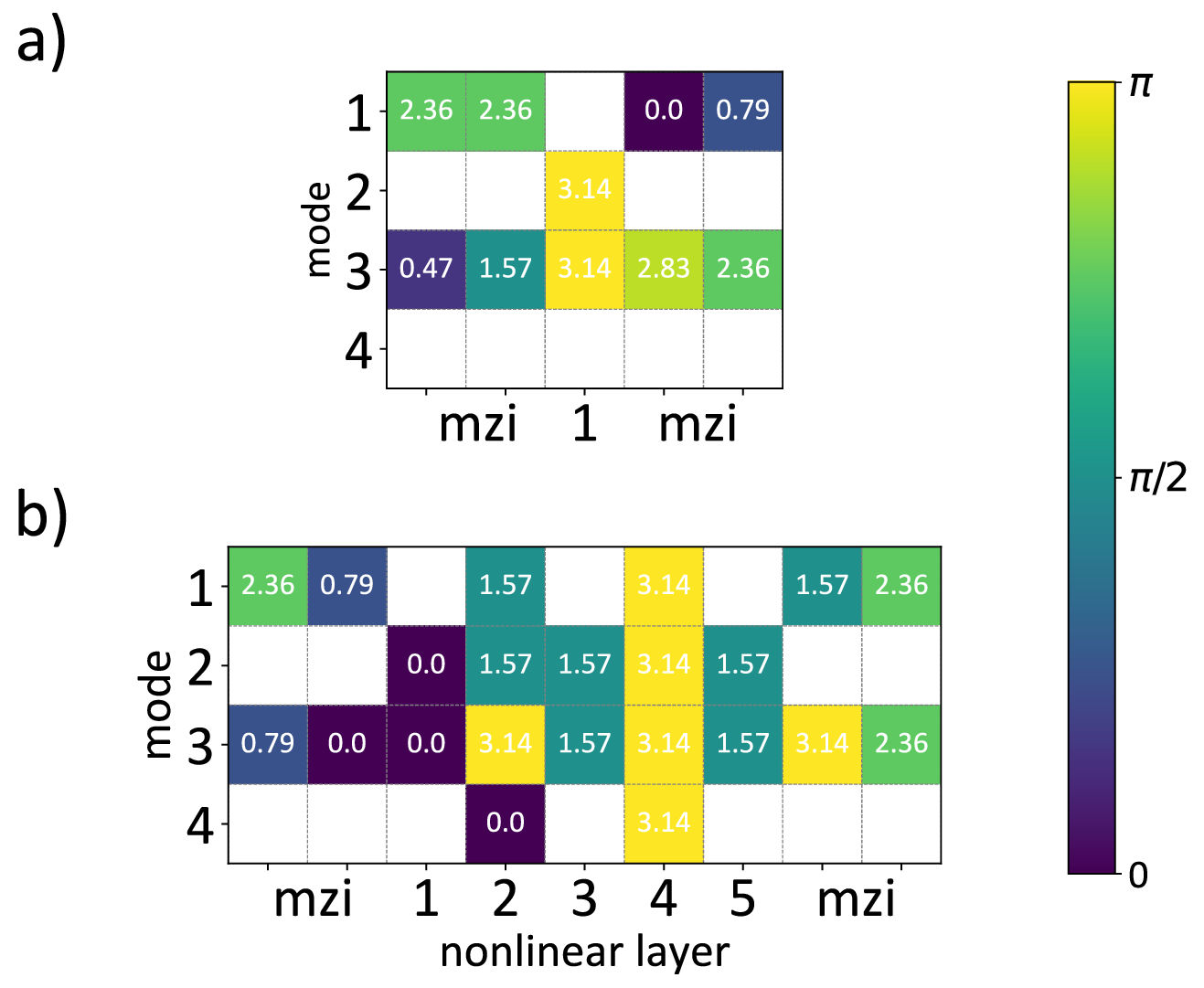}
        \caption{\textbf{Distribution of the parameters of the trained QONNs:} values of the NMZI nonlinearities (designated by the nonlinear layer index) and the MZI phase-shifts (designated by "mzi" label) of the four-state (a) and six-state (b) discriminator at $\varphi_b=0$ with single-qubit corrections. The parameters correspond to the QONNs with  $D_{\text{NL}}=1$ (a) and  $D_{\text{NL}}=5$ (b) (see Fig.~\ref{fig:discriminators_infidelity}).}\label{fig:discriminators_nonlinearities}
    \end{figure}

Fig.~\ref{fig:discriminators_infidelity} shows the dependence of achievable infidelity $1-CF_{disc}$ on the circuit depth $D$ for the four-state (a) and six-state discriminator (b). By default, the variational core of the QONN was trained together with the single-qubit corrections. Additionally, to show the effect of the single-qubit corrections, we separately trained the variational core without the single-qubit corrections. Several values of the NMZI phase-shift biases $\varphi_b$ were considered to investigate its effect and to determine the optimal value.

The results suggest that the NMZI meshes, both with and without single-qubit corrections, can perform state discrimination provided proper phase-shift biases are set. The lowest depths at which the QONNs implement perfect discriminators are achieved at $\varphi_b=0$, corresponding to the identity operation of the NMZI in the linear regime. Specifically, depths of  $D_{\text{NL}}=1$ and  $D_{\text{NL}}=5$ are sufficient for the four- and six-state discrimination, respectively (with single-qubit corrections). A phase-shift of $\varphi_b=\pi$ (corresponding to the mode-swapping operation in the linear regime of NMZI) also works but requires much larger depths:  $D_{\text{NL}}=3$ and  $D_{\text{NL}}=11$ for the four- and six-state discrimination, respectively.

Training only the variational core without the input and output MZIs works as well, however, this comes at increasing the depths even further. In Fig.~\ref{fig:discriminators_infidelity} we plotted the infidelities without the single-qubit corrections for $\varphi_b=0$ and $\varphi_b=\pi$. For these biases, it is possible to achieve convergence of the optimization algorithm only for $\varphi_b=\pi$, because NMZI mode swapping operation allows to redistribute the photons between modes. The minimal circuit depths in this case are  $D_{\text{NL}}=19$ and  $D_{\text{NL}}=23$.

From this example, it is clear that static linear optics significantly impact the capabilities of the QONNs. While NMZI biases $\varphi_b=0$ and $\varphi_b=\pi$ allow our QONN to function as state discriminators, other values considered ($\varphi_b=\pi/4$, $\pi/2$ and $3\pi/4$) do not work. To gain deeper insight into how the QONN discriminates quantum states, we plotted the nonlinearities and the single-qubit correction phase shifts corresponding to the obtained optimal circuits in Fig.~\ref{fig:discriminators_nonlinearities}. The results show that the optimized nonlinear variational cores consist of two types of nonlinearity gates, $\text{NS}(\chi)$, with $\chi=\pi/2$ and $\chi=\pi$. This is due to the inherent symmetry of the task the programmable circuits need to perform.

We summarize the relevant parameters of the QONN performing the state discrimination task in Table~\ref{tab:descriminator}. Notably, our QONN with adjustable nonlinearities is less demanding than QONNs with static nonlinearities, as evident from Fig.~\ref{fig:discriminators_infidelity}. Specifically, the QONN trained with linear optics requires  $D_{\text{LO}}=10$ and $D_{\text{LO}}=20$  programmable linear phase-shift layers to achieve high-quality discrimination of the four- and six-state systems, respectively. Our QONN can do the task using only $D_{\text{NL}}=1$ and $D_{\text{NL}}=5$ programmable nonlinear phase-shift layers (at $\varphi_b=0$). Correspondingly, our QONN requires fewer adjustable nonlinearities than its counterpart trained with linear optics to find the optimal circuits.

\begin{table}
\centering
\caption{Nonlinear $D_{\text{NL}}$ and linear circuit depth  $D_{\text{LO}}$ and parameter count $P$ of the QONNs peforming Bell-state discrimination (from the results shown in Fig.~\ref{fig:discriminators_infidelity}). }
\begin{tblr}{
  cell{1}{1} = {r=2}{},
  vlines,
}
 \hline
 &  $D_{\text{NL}}$ &  $D_{\text{LO}}$ & $P$ \\
\hline
 &   $4$-/$6$-state &  $4$-/$6$-state & $4$-/$6$-state  \\
\hline
NL ($\varphi_b=0$)  & 1/5 &  0 & 2/14 \\
NL ($\varphi_b=\pi$) & 3/11 &  0 & 8/32 \\
NL ($\varphi_b=\pi$,no MZIs) & 19/23 &  0 & 56/68  \\
LO & 1/3 &  10/20 & 24/48  \\
\hline
\end{tblr}\label{tab:descriminator}
\end{table}

\subsection{Variational quantum algorithms}

We next consider QONN performing a variational quantum algorithm (VQA). VQAs enable solving certain types of computational problems by mapping them into corresponding Hamiltonians $\hat{H}$, whose ground state encodes the solutions. In order to find a ground state by VQA, one prepares parametrized trial states $|\Psi(\boldsymbol{\chi},\boldsymbol{\theta})\rangle$ on a quantum computer and estimates the Hamiltonian expectation value by measuring $E(\boldsymbol{\chi},\boldsymbol{\theta})=\langle\psi^{(dr)}(\boldsymbol{\chi},\boldsymbol{\theta})|\hat{H}|\psi^{(dr)}(\boldsymbol{\chi},\boldsymbol{\theta})\rangle$ --- the cost function to be optimized by the classical computer, the global minimum $E_{\text{min}}\le{}E(\boldsymbol{\chi},\boldsymbol{\theta})$ corresponds to a solution. Here, $|\psi^{(dr)}(\boldsymbol{\chi},\boldsymbol{\theta})\rangle$ designates the dual rail part of the output state $|\Psi(\boldsymbol{\chi},\boldsymbol{\theta})\rangle$. In contrast to implementing specific quantum operations, such as those described above, solving problems with the variational quantum eigensolver (VQE) does not require the exact generation of a specified state. VQE may provide good approximate solutions for a class of prepared quantum states that are not necessarily coinciding with the true ground state. Despite this relaxed complexity, solving practical problems by VQA is coupled with preparation of large-qubit quantum states, which, on the one hand, is too sophisticated due to the limitations of the current NISQ devices and, on the other hand, hard to optimize because of the unsolved issue of barrens plateau~\cite{BarrenPlateaus}. Therefore, developing less demanding quantum architectures for VQE is required.

In this section, we investigate our QONN performing variational search of  the ground state energy of the Heisenberg Hamiltonians. In general, the Heisenberg Hamiltonian takes the form:
    \begin{equation}\label{eqn:Heisenberg_hamiltonian}
    \begin{split}
        H_{\text{Heis}} = -\sum_{<i,j>\in{}G^{(x)}} J_{ij}^{(x)} \hat{X}_i
        \hat{X}_j - \sum_{<i,j>\in{}G^{(y)}} J_{ij}^{(z)} \hat{Y}_i
        \hat{Y}_j - \\
        -  \sum_{<i,j>\in{}G^{(z)}} J_{ij}^{(z)} \hat{Z}_i 
        \hat{Z}_j - \sum_i h_i \hat{X}_i
    \end{split}
    \end{equation}
where $\hat{X}_i, \hat{Y}_i, \hat{Z}_i$ are the Pauli operators describing spin with index $i$. A particular Hamiltonian is specified by the interactions between the spins described by interaction coefficients $J_{ij}^{(\mu)}$ ($\mu=x,y,z$) with interaction graphs $G^{(\mu)}$ and external fields $h_i$. Given specified interactions and external fields, VQA boils down to finding the Hamiltonian ground state energy using an optimization algorithm run on a classical computer. Depending on the interaction graphs of the Hamiltonians, various quantum systems with different properties can be catched~\cite{zhong2002theoretical,banchi2017pretty}

\begin{figure}[htp]
        \centering\includegraphics[width=0.4\textwidth]{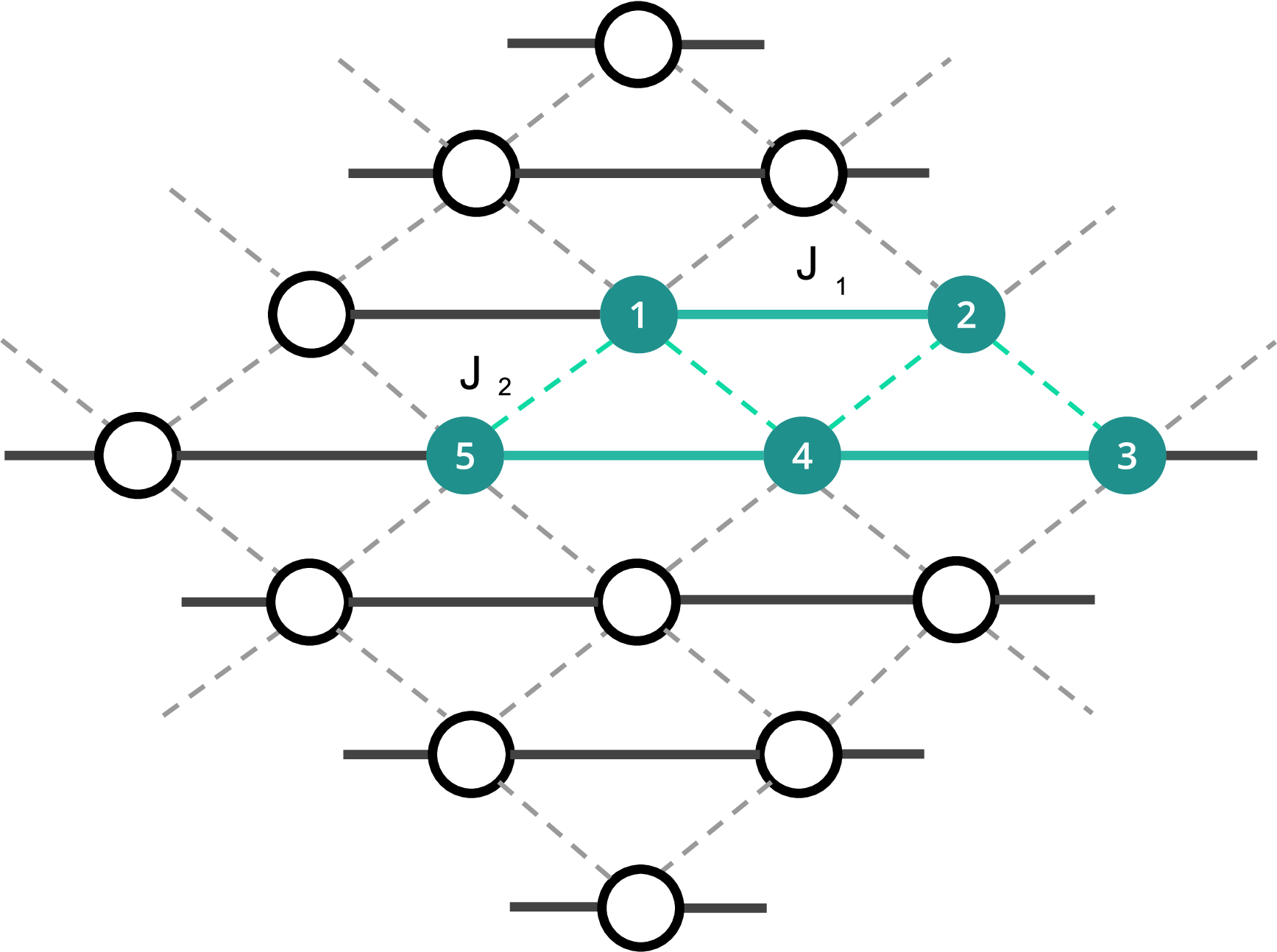}
        \caption{\textbf{The 2D triangular Heisenberg lattice model. } The lattice is linear chains of spins with the interaction described by parameters $J_1$ and $J_2$ corresponding to intra- and interchain coupling strengths, respectively. The highlighted $5$-qubit graph is the topology of the studied Hamiltonian in the QONN VQA simulation (the graph is assumed uncoupled from the outer nodes).     }\label{fig:triangular_lattice}
    \end{figure}

As a specific example, we consider a 2D triangular lattice shown in Fig.~\ref{fig:triangular_lattice}, which occurs in some materials~\cite{weihong1999phase, zheng2006excitation}. The Hamiltonian displays nearest-neighbour interactions characterized by coefficients $J_{i,\text{nbr}(i)}=J_{i,\text{nbr}(i)}^{(x)}=J_{i,\text{nbr}(i)}^{(y)}=J_{i,\text{nbr}(i)}^{(z)}$. The lattice is a set of linear chains of spins with two  coefficients, $J_1$ and $J_2$, that describe intra- and inter-chain couplings,respectively. 

    \begin{figure}[htbp]
        \centering\includegraphics[width=0.48\textwidth]{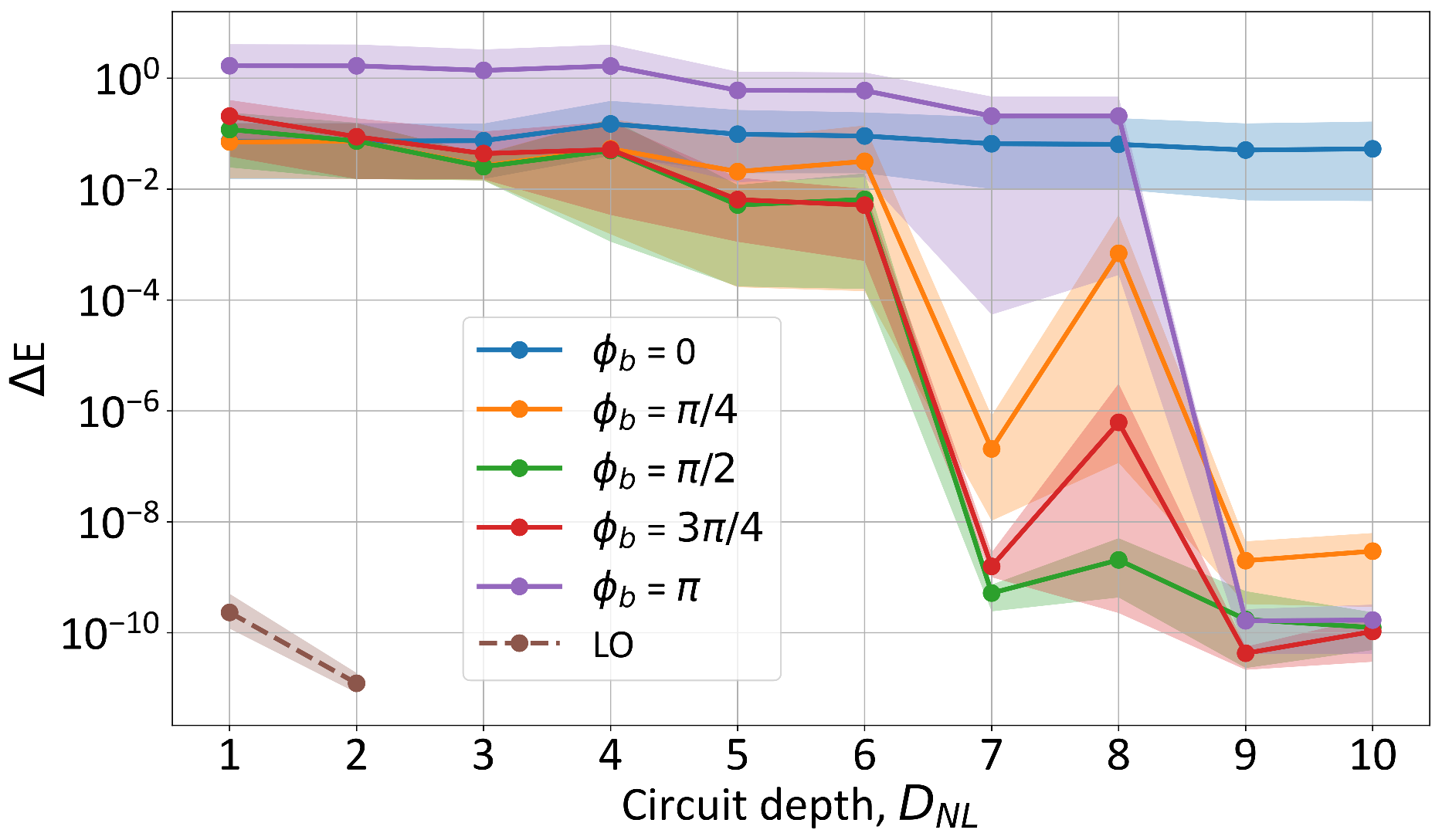}
        \caption{\textbf{VQE of the Heisenberg Hamiltonian performed by the QONNs}: energy error $\Delta{}E=E_{\text{exact}}-E_{\text{qonn}}$ as a function of depth  $D_{\text{NL}}$ of the QONN programmed by nonlinearities at different values the NMZI phase-shift bias $\varphi_b$ (solid curves) and the dependence for the QONN programmed by linear optics (dashed curve). Each dependence correspond to $15$ Hamiltonians, obtained by randomly generating  $J_1$, $J_2$ and $h_i$ from a uniform distribution in the range from $0$ to $1$.  For each value of $D_{\text{NL}}$ and each random Hamiltonian, the best $\Delta{}E$ from $50$ optimization runs was taken into the distribution.}\label{fig:vqe5_Heisenberg}
    \end{figure}

Specifically, we sampled a number of Hamiltonians by generating values of $J_1$, $J_2$ and $h_i$ ($i=\overline{1,5}$) randomly from the uniform distribution over in the range $[0,1]$. Here, we quantify the infidelity by the energy error $\Delta{}E=E_{\text{exact}}-E_{\text{qonn}}$, where $E_{\text{qonn}}$ is the energy obtained by training/optimizing a QONN, $E_{\text{exact}}$ is its exact value obtained by the diagonalization of the Hamiltonian matrix using a standard library (Python SciPy). The obtained dependencies of error on the circuit depth  $D_{\text{NL}}$ is shown in Fig.~\ref{fig:vqe5_Heisenberg}. As before, not all static phase-shift values $\varphi_b$ enabled find as a high-accuracy solution. As we see, the value $\varphi_b=0$ (corresponding to the mode swapping operation in the linear regime of NMZI) does not yield satisfactory quality. At the same time, all the rest of the values work almost equally good.

According to Fig.~\ref{fig:vqe5_Heisenberg}, it requires minimum $D_{\text{NL}}=7$ programmable nonlinear layers for the best case of $\varphi_b=\pi/2$ for our QONN to be able to find high-quality  solutions, while the QONN programmed by linear optics requires $D_{\text{LO}}=22$ programmable linear layers. This suggests that our QONN is advantageous.

\begin{table}
\centering
\caption{Nonlinear $D_{\text{NL}}$ and linear circuit depth  $D_{\text{LO}}$ and parameter count $P$ in the QONNs peforming VQE of the Heisenberg Hamiltonian (see Fig.~\ref{fig:vqe5_Heisenberg}). NL and LO are the QONNs with adjustable nonlinearities and the counterparts trained by linear optics, respectively.}
\begin{tblr}{
  vlines,
}
\hline
  & $D_{\text{NL}}$ & $D_{\text{LO}}$ & $P$  \\
\hline
NL ($\varphi_b=\pi/2$) & 7 & 0 & 62   \\
LO  & 1 &  22 & 180  \\
\hline
\end{tblr}\label{tab:vqe5_heis_table}
\end{table}

It is worth noting that the circuit depth, and therefore the number of adjustable parameters in QONNs, can be further reduced by using different building blocks. Specifically, the ability of a QONN layer to implement non-local interactions may be more advantageous for certain problems compared to the two-mode interactions of NMZI layers. Our proposal is not confined to this specific QONN architecture. We suggest exploring the use of programmable nonlinearities in other architectures that may be developed in the future.

\subsection{Training complexity}

In addition to developing improved hardware architectures for quantum neural networks that optimize the use of available physical resources, their trainability is also of critical importance. The challenge of training parametrized quantum circuits, primarily due to the presence of Barren Plateaus in the loss function~\cite{BarrenPlateausReview}, presents a significant obstacle to leveraging variational quantum computing for solving practical tasks. The optical implementations of quantum neural networks are expected to exhibit the manifestation of Barren Plateaus as well. However, the peculiarities in the "hardness" of the loss functions to be optimized in the neural networks requires special analysis, which will be pursued separately.

    \begin{figure}[htbp]
        \centering\includegraphics[width=0.50\textwidth]{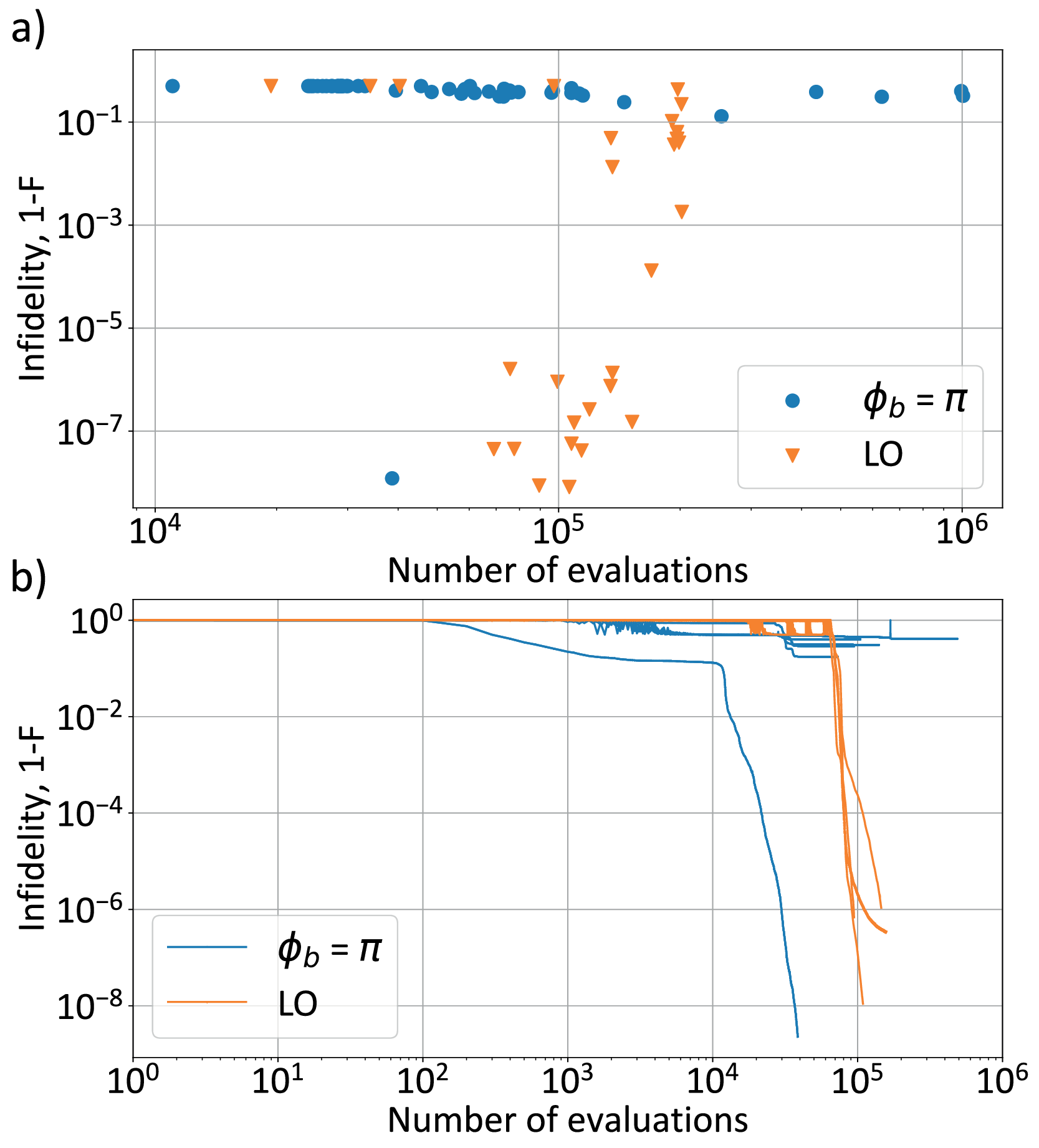}
        \caption{\textbf{Training the QONNs for preparation of the $4$-qubit GHZ-state}: a) the set of infidelities to which the optimization algorithm has converged in $30$ optimization runs and the corresponding number of evaluations of the cost function; b) examples of convergence curves in $10$ optimization runs. The parameters of the QONNs correspond to Fig.~\ref{fig:ghz4_gen}e) and Fig.~\ref{fig:ghz4_gen}f) at $D_{\text{NL}}=9$ and $D_{\text{NL}}=3$, respectively.  }\label{fig:training_complexity}
    \end{figure}

In this section, we compare two types of QONNs: one based on NMZI, trained by adjusting nonlinearities, and another trained by programmable linear optics. Typically, a single optimization run does not guarantee convergence to the global minimum of the cost function. Therefore, multiple independent runs were conducted for each target state or transformation to achieve a satisfactory solution. As a hint on the trainability of our NMZI-based QONN, let us consider the task of entangled states generation, studied above. As a concrete example, Fig.~\ref{fig:training_complexity} compares the training complexity of the QONNs targeted at generation of the $4$-qubit GHZ-state. Namely, Fig.~\ref{fig:training_complexity}a shows the distribution of cost function values obtained in $30$ optimization runs. Fig.~\ref{fig:training_complexity}b presents examples of convergence curves. The depths of both QONNs were set to achieve high-fidelity optima, as shown in Fig.~\ref{fig:ghz4_gen}e) and Fig.~\ref{fig:ghz4_gen}f) for our NMZI-based QONN and for the QONN programmed by linear optics, respectively. As can be seen from Fig.~\ref{fig:training_complexity}a, only a fraction of the obtained values meets the high-fidelity requirements (say $1-F\le{}10^{-7}$).

From Fig.~\ref{fig:training_complexity}a, we observe that the network trained with linear optics demonstrates faster convergence and requires fewer optimization runs than the NMZI-based network trained by adjusting nonlinearities. This trend is consistent across other types of problems considered above. However, we cannot conclude that this is an inherent characteristic of QONNs trained by nonlinearities. Instead, this finding suggests the need for exploring alternative optimization algorithms and potentially developing new ones tailored to the specific architecture.


\section{Discussion and Outlook}\label{sec:discussion}

We have proposed a novel approach for constructing parametrized quantum optical circuits using programmable nonlinearities. Using NMZI meshes as an example of such programmable circuits, we have demonstrated that high-quality operation can be achieved by training these nonlinearities. The key insight of our work is that relying solely on programmable linear optical interference is suboptimal for implementing QONNs.  Our approach, which incorporates programmable nonlinearities, significantly reduces the number of layers of programmable elements, thereby decreasing errors from circuit components and resulting in more compact designs. Additionally, variational circuits can benefit from a lower number of adjustable parameters, allowing for more computationally efficient optimization algorithms, although further studies are required. Although our analysis focused on a specific architecture, we expect the results to be more broadly applicable. Other architectures using programmable nonlinearities may also see even more pronounced improvements. Also, extending our approach to the continuous variable quantum neural networks~\cite{CV_NN} may yield similar improvements.

We outline several open questions for future investigation. An important question involves the constituent optical blocks used to construct programmable quantum optical circuits. Specifically, it is of interest to explore how the choice of linear and nonlinear optical components affects the performance of QONNs. The NMZI block used in this work consists of two-mode linear optical DCs, but a variety of multimode couplers could be used to design other types of building blocks. Additionally, while we focused on Kerr-like nonlinearity, there are many other nonlinearities that may act in more complex ways, including various multimode nonlinearities. These potentially accessible nonlinearities together with multimode linear optical elements could facilitate the construction of more expressive and/or more resource-efficient quantum optical circuits by providing more sophisticated multivariate elements.

Another question pertains to methods for designing variational ansatze tailored to specific classes of problems, similar to gate-based variational circuits. In gate-based implementations of variational quantum algorithms, effectively solving a problem often requires designing a specific circuit architecture that accounts for problem specifics such as symmetries or conservation laws. This could also apply to the quantum optical platform, although methods from gate-based models cannot be directly used. Therefore, research in this direction is necessary.

In conclusion, we cannot help but point out the analogy with the field digital neural networks occurs recently with the proposal of Kolmogorov-Arnold networks~\cite{KAN_NN}. KANs rely on training nonlinear activation functions, in contrast to training linear weights in the traditional Multi-Layer Perceptrons architecture. According to~\cite{KAN_NN}, this can yield neural networks with smaller parameter space, but more challenging process of training the activation functions. Our results suggest that the advantages of the QONNs with trained nonlinearities can come at the cost of using more sophisticated training algorithms.

\section{Acknowledgements}

The authors acknowledge support by Rosatom in the framework of the Roadmap for Quantum computing (Contract No. 868-1.3-15/15-2021 dated October 5, 2021 and Contract No. P2154 dated November 24, 2021). E.A.C is  grateful to the Foundation for the Advancement of Theoretical Physics and Mathematics (BASIS) (Project No. 23-2-1-52-1). M.Yu.S and S.P.K. acknowledge partial support by Ministry of Science and Higher Education of the Russian Federation and South Ural State University (agreement No. 075-15-2022-1116).

\appendix

\section{Transformation of the NMZI block}\label{sec:nmzi_transformation}

We consider a two-mode block, shown in Fig.~\ref{fig:qonn_architecture}c which we call nonlinear MZI (NMZI). The NMZIs are constructed from the  balanced directional couplers (DCs) characterized by the transfer matrices $U_{\text{DC}}=\frac{1}{\sqrt{2}}\left(\begin{array}{cc}
    1 & 1 \\
    1 & -1
\end{array}\right)$, and two programmable nonlinear sign-shift gates (NS-gates) placed in the arms between the DCs. The equation for NMZI contained two nonlinear elements $\hat{\text{NS}}_m(\chi)=\exp\left(i\frac{\chi}{2}(\hat{a}_m^{\dagger})^2\hat{a}_m^2\right)$~\eqref{eqn:nonlinearity} with parameters $\chi_1$ and $\chi_2$ and linear phase shifter $\varphi$ for any input Fock state $ |\psi_{in}\rangle = |n,m\rangle$ can be written as
\begin{equation}
    \begin{split}\label{NMZI_formula}
        |\psi_{out}\rangle = \hat{U}_{\text{NMZI}}|n,m\rangle = \left(\dfrac{1}{2}\right)^{n+m}\sum_{p=0}^{n+m}\sum_{j=0}^p \begin{pmatrix} n\\ j\end{pmatrix} 
        \times \\
        \begin{pmatrix} m\\ p-j \end{pmatrix}
        (-1)^{m-p+j} \times \\ \times
        e^{i p(p-1)\frac{\chi_1}{2}} 
        e^{i q(q-1)\frac{\chi_2}{2}}  e^{i\varphi p} 
        \times \\
        \sum_{\Tilde{p}=0}^{n+m}\sum_{k=0}^{\Tilde{p}} \begin{pmatrix} p\\ k\end{pmatrix} \begin{pmatrix} q\\ \Tilde{p}-k\end{pmatrix}  
        (-1)^{q-\Tilde{p}+k} \dfrac{\sqrt{\Tilde{p}!\Tilde{q}!}}{\sqrt{n!m!}}| \Tilde{p},\Tilde{q} \rangle,
    \end{split}
\end{equation}
where $q=n+m-p$, $\Tilde{q} = n+m-\Tilde{p}$, $\begin{pmatrix} n\\ j\end{pmatrix} = C_n^j$, $\begin{pmatrix} m\\ p-j\end{pmatrix} = C_m^{p-j}$, $\begin{pmatrix} p\\ k\end{pmatrix} = C_p^k$, $\begin{pmatrix} q\\ \Tilde{p}-k\end{pmatrix} = C_q^{\Tilde{p}-k}$ are binomial coefficients.

    \begin{figure}[htp]
        \centering\includegraphics[width=0.47\textwidth]{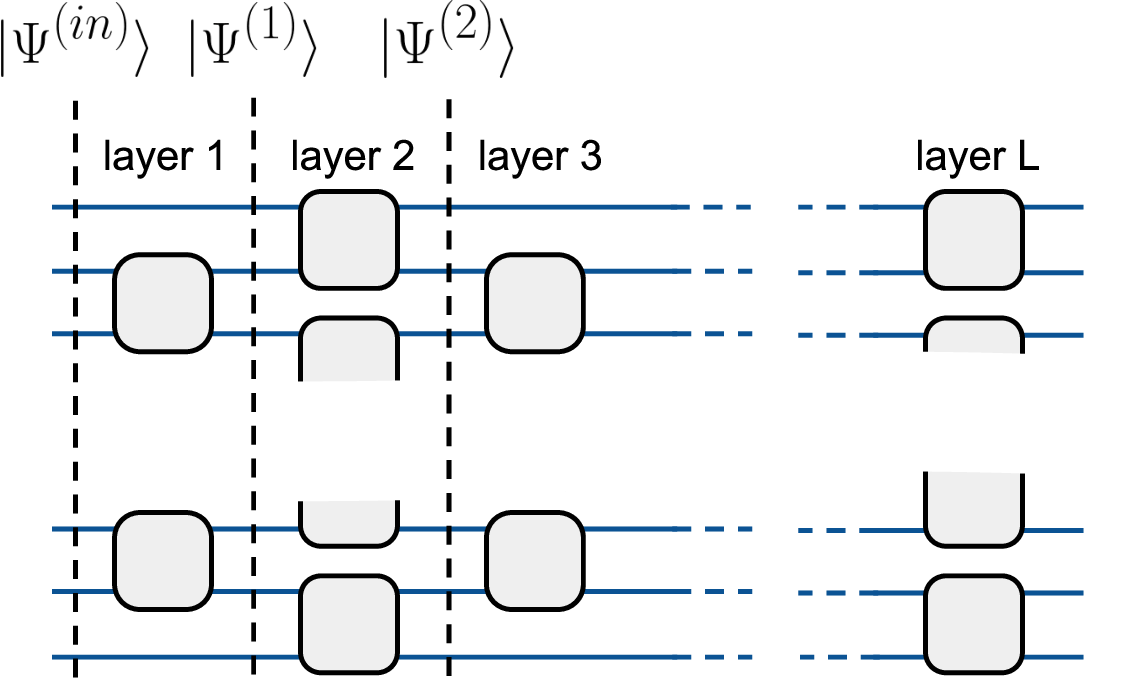}
        \caption{\textbf{Transformation of the state vector in the QONN simulation on a classical computer}: input state $|\Psi^{(in)}\rangle$ change after sequentially traversing the layers of the neural network;
        }\label{fig:NMZI_transformation}
    \end{figure}

The QONN simulation operation on a classical computer executes a layer-by-layer input state $|\Psi^{(in)}\rangle$ transformation, as shown in the Fig.~\ref{fig:NMZI_transformation}. In general terms quantum state after layer $k$ can be represented in the form  $|\Psi^{(l)}\rangle = \sum_{\boldsymbol{t}\in\Omega_N^M}c^{(l)}_{\boldsymbol{t}}|\boldsymbol{t}\rangle$, where $\boldsymbol{t}=(t_1,t_2,\ldots,t_M)$ is the photon-occupation vector designating a particular $N$-photon Fock state $|\boldsymbol{t}\rangle$ ($\sum_{m=1}^M{}t_m=N$) and the corresponding probability amplitude $c^{(l)}_{\boldsymbol{t}}$, $\Omega_N^M$ is the set of all possible $N$-photon $M$-mode Fock states. State $|\Psi^{(l-1)}\rangle = |\Psi^{(l)}_0\rangle$ transforms into state $|\Psi^{(l)}\rangle$ by consistently applying $\hat{U}_{\text{NMZI},i}$ (where $i = \overline{1, \lfloor M/2 \rfloor}$, if $l$ is even, and $i = \overline{1, \lfloor M/2 \rfloor-1}$, if $l$ is odd) between adjacent modes according to their location in the circuit. For example, calculating the $i$-th NMZI transformation ($l$ is an even number): 
\begin{equation}
    \begin{split}\label{QONN_transform}
    |\Psi^{(l)}_{i}\rangle = \hat{U}_{\text{NMZI},i}|\Psi^{(l)}_{i-1}\rangle =  \sum_{\boldsymbol{t}\in\Omega_N^M}c^{(l)}_{\boldsymbol{t},i-1} \times \\ 
    \times |t_{1}, ..., t_{2i-2}\rangle \otimes \hat{U}_{\text{NMZI}}|t_{2i-1},t_{2i}\rangle \otimes |t_{2i+1},...,t_{M}\rangle
    \end{split}
\end{equation}

\section{Optimization algorithm}\label{sec:optimization}

QONN training runs in two consecutive stages: preliminary global optimization giving a coarse solution followed by the local one, which improves the solution. The global optimizer is run with a low target precision --- the iterations are stopped when the change of the cost function in one iteration becomes smaller than $10^{-4}$. This allows to find a possible extremum basin and carefully choose a starting point for subsequent local optimization. The local optimizer refines the solution and stops when cost function accuracy $10^{-14}$ is reached. Therefore, under random initial conditions, such pair of optimizers do not converge to the nearest minimum and explore the parameter space more thoroughly.

We use gradient-free optimization algorithms that optimize an objective function without explicitly defined derivatives. We utilized Controlled Random Search (CRS)~\cite{kaelo2006some} as the global optimizer and Bound Optimization BY Quadratic Approximation (BOBYQA)~\cite{powell2009bobyqa} for local optimization as implemented in NLopt library~\cite{johnson2014nlopt}. CRS algorithm starts with a random population of points with size $N = 10 \times (P+1)$, and evolves these points by specified rules. It was chosen because of its ability to work efficiently in large-dimensional parameter spaces, which is difficult for many optimization algorithms. The size of the parameter space plays an important role in training a deep neural network. For the same reason, BOBYQA is preferred. This algorithm performs bound-constrained optimization using an iteratively constructed quadratic approximation for the objective function using $m$ interpolation points, the value $m = 2P+1$ being typical. Moreover, the combination of these two algorithms demonstrated the best convergence among other methods from the NLopt library that we have tested.

\begin{figure}[htp]
        \centering\includegraphics[width=0.45\textwidth]{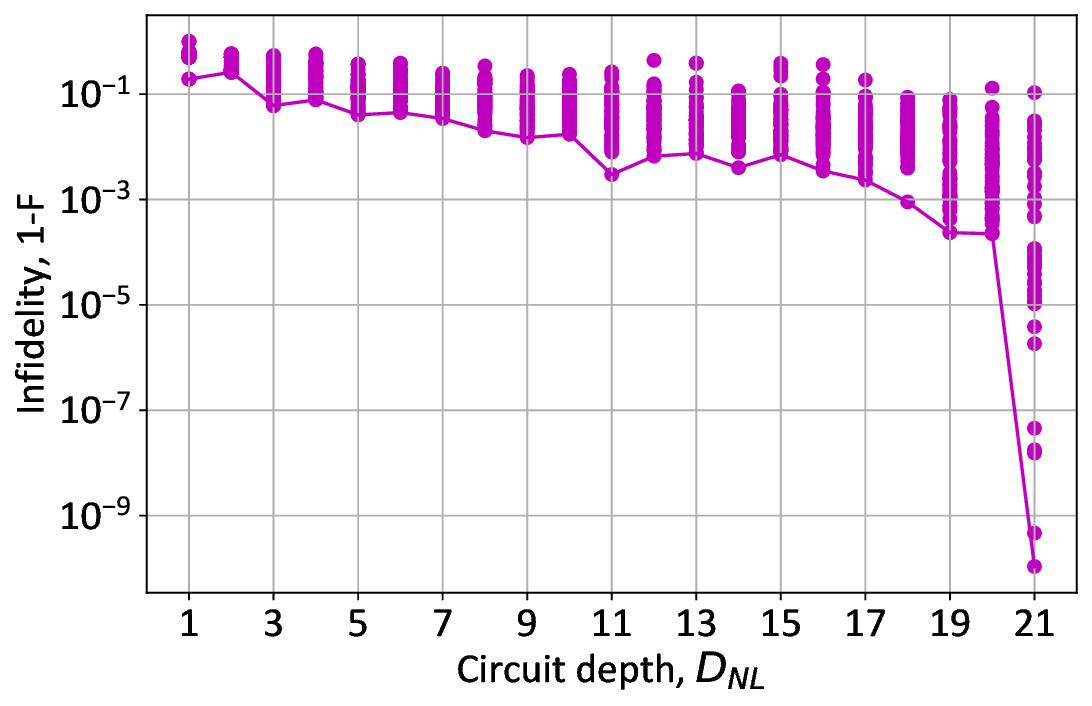}
        \caption{ Example of infidelity values obtained in multiple runs of optimizatin algorithm for the QONN generating $3$-qubit state $|G_N(\alpha)\rangle$ at $\alpha=\pi/4$ and $\varphi_b=\pi/4$. The best values (marked by connected points) to one curve depicted in Fig.~4b. }
        \label{fig:infidelity_distribution}
    \end{figure}

It is necessary to run the optimization algorithm multiple times to achieve a minimum value of the cost function. To illustrate this, Fig.~\ref{fig:infidelity_distribution} shows the dependence of infidelity for the QONN with adjustable nonlinearities that prepares the $3$-qubit state $|G_N(\alpha)\rangle$ for $\alpha=\pi/4$. At each value depth $D_{\text{NL}}$ the optimization algorithm was ran $50$ times, each time initialized by random values of the QONN parameters. The points in the figure correspond to the infidelities obtained. The best obtained values (connected points)  corresponds to one dependence from Fig. 4b.

\section{Multi-photon dynamics inside the QONNs}\label{sec:multiphoton}

    \begin{figure*}[htp]
        \centering\includegraphics[width=0.78\textwidth]{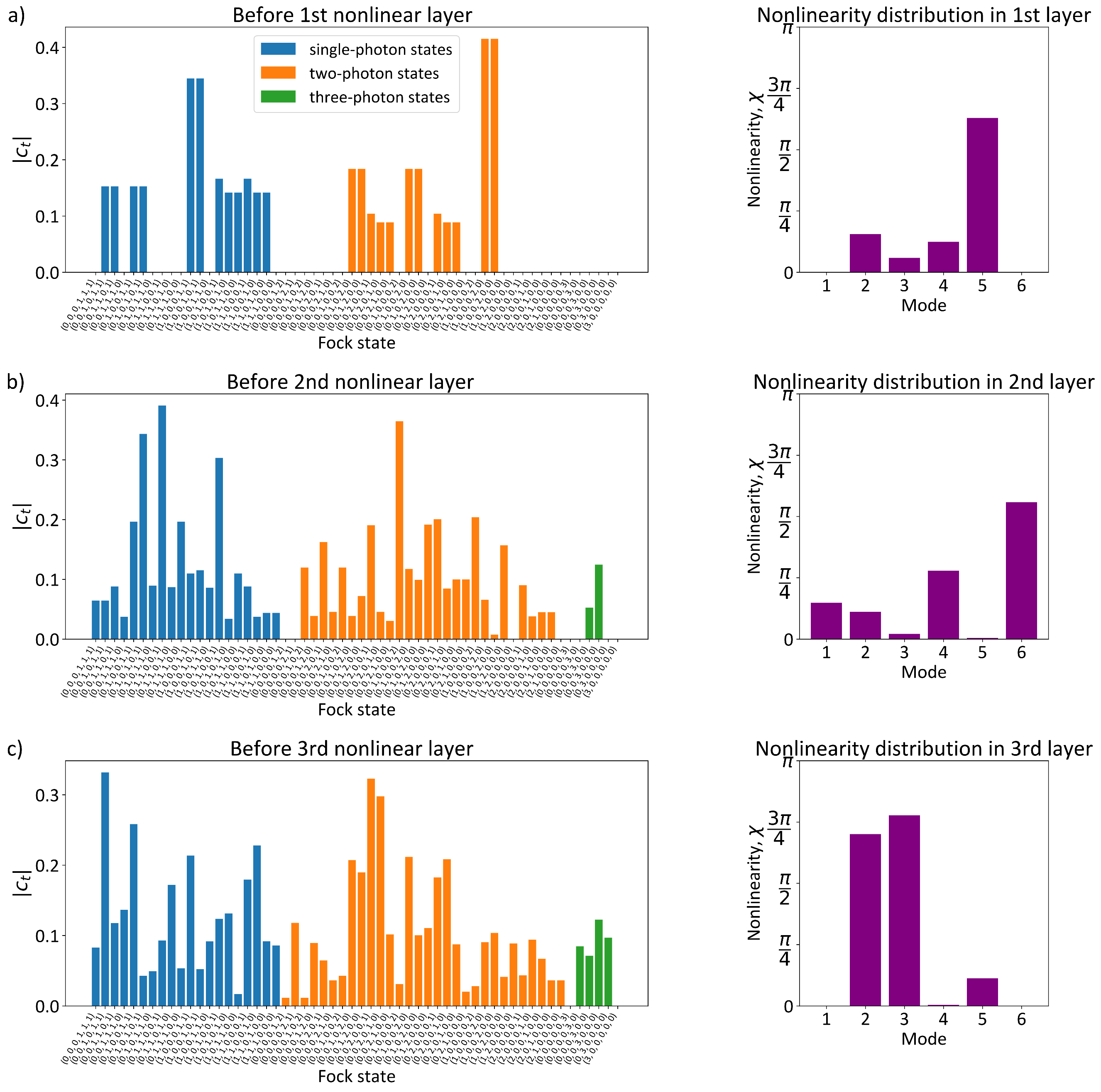}
        \caption{(left column) Example of distributions of absolute values of the probability amplitudes of the quantum states entering the nonlinear layers of the QONN with adjustable nonlinearities and (right column) corresponding nonlinearity strengths within the quantum circuit. The QONN has been trained to perform $3$-qubit VQE algorithm with Heisenberg Hamiltonian on the $2D$ triangular lattice model. The QONN has $D_{\text{NL}}=3$ programmable nonlinear layers.}
        \label{fig:VQE3_ampl_distribution}
    \end{figure*}
    \begin{figure*}[htp]
        \centering\includegraphics[width=0.62\textwidth]{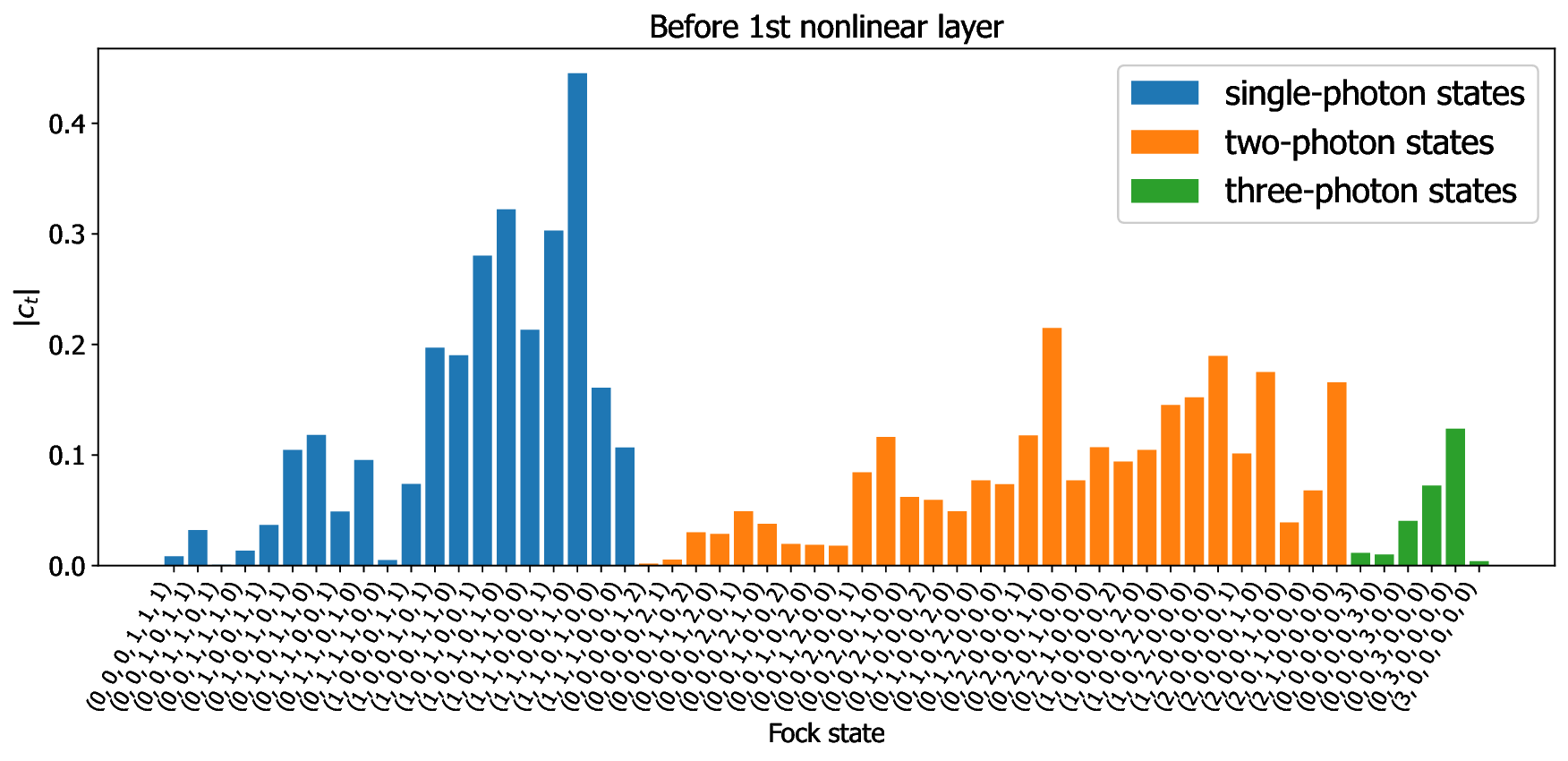}
        \caption{ Example of distributions of absolute values of the probability amplitudes of the quantum states entering the nonlinear layers of the QONN with programmable linear optics. The QONN has been trained to perform $3$-qubit VQE algorithm with Heisenberg Hamiltonian on the $2D$ triangular lattice model. The QONN has $D_{\text{NL}}=1$ nonlinear layer with static nonlinearities $\chi = \pi$.}
        \label{fig:VQE3_LO_ampl_distribution}
    \end{figure*}

The QONNs operates in the Hilbert space ${\cal{}H}_N^M$ of discrete $N$-photon states occupying $M$-modes, in which the states are superpositions of the form $\sum_{\boldsymbol{t}\in\Omega_N^M}c_{\boldsymbol{t}}|\boldsymbol{t}\rangle$, where $|\boldsymbol{t}\rangle$ is the $N$-photon Fock state, $\Omega_N^M$ is the set of all possible $N$-photon $M$-mode Fock states. In this paper, we prepare the input qubit states in dual-rail encoding, but when photons pass through the mesh of two-mode NMZIs, multi-photon states inevitably arise within the linear-optical circuit. Nonlinear elements $\hat{\text{NS}}_m(\chi)=\exp\left(i\frac{\chi}{2}(\hat{a}_m^{\dagger})^2\hat{a}_m^2\right)$ are not effective for single-photon components of the states, so in this section we demonstrate that the probability amplitudes of states within the QONN before nonlinearities take significant values for multi-photon components.

Fig.~\ref{fig:VQE3_ampl_distribution} and fig.~\ref{fig:VQE3_LO_ampl_distribution} show an example of the probability amplitude distribution of the state before the nonlinear layers within neural network with programmable nonlinearities and linear optical network, respectively, for the VQE of the 3-qubit Heisenberg Hamiltonian $H$ on the 2D triangular lattice model.

\newpage

\bibliography{apssamp}

\begin{thebibliography}{56}%
\makeatletter
\providecommand \@ifxundefined [1]{%
 \@ifx{#1\undefined}
}%
\providecommand \@ifnum [1]{%
 \ifnum #1\expandafter \@firstoftwo
 \else \expandafter \@secondoftwo
 \fi
}%
\providecommand \@ifx [1]{%
 \ifx #1\expandafter \@firstoftwo
 \else \expandafter \@secondoftwo
 \fi
}%
\providecommand \natexlab [1]{#1}%
\providecommand \enquote  [1]{``#1''}%
\providecommand \bibnamefont  [1]{#1}%
\providecommand \bibfnamefont [1]{#1}%
\providecommand \citenamefont [1]{#1}%
\providecommand \href@noop [0]{\@secondoftwo}%
\providecommand \href [0]{\begingroup \@sanitize@url \@href}%
\providecommand \@href[1]{\@@startlink{#1}\@@href}%
\providecommand \@@href[1]{\endgroup#1\@@endlink}%
\providecommand \@sanitize@url [0]{\catcode `\\12\catcode `\$12\catcode `\&12\catcode `\#12\catcode `\^12\catcode `\_12\catcode `\%12\relax}%
\providecommand \@@startlink[1]{}%
\providecommand \@@endlink[0]{}%
\providecommand \url  [0]{\begingroup\@sanitize@url \@url }%
\providecommand \@url [1]{\endgroup\@href {#1}{\urlprefix }}%
\providecommand \urlprefix  [0]{URL }%
\providecommand \Eprint [0]{\href }%
\providecommand \doibase [0]{https://doi.org/}%
\providecommand \selectlanguage [0]{\@gobble}%
\providecommand \bibinfo  [0]{\@secondoftwo}%
\providecommand \bibfield  [0]{\@secondoftwo}%
\providecommand \translation [1]{[#1]}%
\providecommand \BibitemOpen [0]{}%
\providecommand \bibitemStop [0]{}%
\providecommand \bibitemNoStop [0]{.\EOS\space}%
\providecommand \EOS [0]{\spacefactor3000\relax}%
\providecommand \BibitemShut  [1]{\csname bibitem#1\endcsname}%
\let\auto@bib@innerbib\@empty
\bibitem [{\citenamefont {Preskill}(2018)}]{PreskillNISQ}%
  \BibitemOpen
  \bibfield  {author} {\bibinfo {author} {\bibfnamefont {J.}~\bibnamefont {Preskill}},\ }\bibfield  {title} {\bibinfo {title} {Quantum {C}omputing in the {NISQ} era and beyond},\ }\href {https://doi.org/10.22331/q-2018-08-06-79} {\bibfield  {journal} {\bibinfo  {journal} {{Quantum}}\ }\textbf {\bibinfo {volume} {2}},\ \bibinfo {pages} {79} (\bibinfo {year} {2018})}\BibitemShut {NoStop}%
\bibitem [{\citenamefont {Cerezo}\ \emph {et~al.}(2022)\citenamefont {Cerezo}, \citenamefont {Verdon}, \citenamefont {Huang}, \citenamefont {Cincio},\ and\ \citenamefont {Coles}}]{CerezoML}%
  \BibitemOpen
  \bibfield  {author} {\bibinfo {author} {\bibfnamefont {M.}~\bibnamefont {Cerezo}}, \bibinfo {author} {\bibfnamefont {G.}~\bibnamefont {Verdon}}, \bibinfo {author} {\bibfnamefont {H.-Y.}\ \bibnamefont {Huang}}, \bibinfo {author} {\bibfnamefont {L.}~\bibnamefont {Cincio}},\ and\ \bibinfo {author} {\bibfnamefont {P.~J.}\ \bibnamefont {Coles}},\ }\bibfield  {title} {\bibinfo {title} {Challenges and opportunities in quantum machine learning},\ }\href {https://doi.org/10.1038/s43588-022-00311-3} {\bibfield  {journal} {\bibinfo  {journal} {Nature Computational Science}\ }\textbf {\bibinfo {volume} {2}},\ \bibinfo {pages} {567} (\bibinfo {year} {2022})}\BibitemShut {NoStop}%
\bibitem [{\citenamefont {Biamonte}\ \emph {et~al.}(2017)\citenamefont {Biamonte}, \citenamefont {Wittek}, \citenamefont {Pancotti}, \citenamefont {Rebentrost}, \citenamefont {Wiebe},\ and\ \citenamefont {Lloyd}}]{BiamonteML}%
  \BibitemOpen
  \bibfield  {author} {\bibinfo {author} {\bibfnamefont {J.}~\bibnamefont {Biamonte}}, \bibinfo {author} {\bibfnamefont {P.}~\bibnamefont {Wittek}}, \bibinfo {author} {\bibfnamefont {N.}~\bibnamefont {Pancotti}}, \bibinfo {author} {\bibfnamefont {P.}~\bibnamefont {Rebentrost}}, \bibinfo {author} {\bibfnamefont {N.}~\bibnamefont {Wiebe}},\ and\ \bibinfo {author} {\bibfnamefont {S.}~\bibnamefont {Lloyd}},\ }\bibfield  {title} {\bibinfo {title} {Quantum machine learning},\ }\href {https://doi.org/10.1038/nature23474} {\bibfield  {journal} {\bibinfo  {journal} {Nature}\ }\textbf {\bibinfo {volume} {549}},\ \bibinfo {pages} {195} (\bibinfo {year} {2017})}\BibitemShut {NoStop}%
\bibitem [{\citenamefont {Huang}\ \emph {et~al.}(2022)\citenamefont {Huang}, \citenamefont {Kueng}, \citenamefont {Torlai}, \citenamefont {Albert},\ and\ \citenamefont {Preskill}}]{PreskillML}%
  \BibitemOpen
  \bibfield  {author} {\bibinfo {author} {\bibfnamefont {H.-Y.}\ \bibnamefont {Huang}}, \bibinfo {author} {\bibfnamefont {R.}~\bibnamefont {Kueng}}, \bibinfo {author} {\bibfnamefont {G.}~\bibnamefont {Torlai}}, \bibinfo {author} {\bibfnamefont {V.~V.}\ \bibnamefont {Albert}},\ and\ \bibinfo {author} {\bibfnamefont {J.}~\bibnamefont {Preskill}},\ }\bibfield  {title} {\bibinfo {title} {Provably efficient machine learning for quantum many-body problems},\ }\href {https://doi.org/10.1126/science.abk3333} {\bibfield  {journal} {\bibinfo  {journal} {Science}\ }\textbf {\bibinfo {volume} {377}},\ \bibinfo {pages} {eabk3333} (\bibinfo {year} {2022})},\ \Eprint {https://arxiv.org/abs/https://www.science.org/doi/pdf/10.1126/science.abk3333} {https://www.science.org/doi/pdf/10.1126/science.abk3333} \BibitemShut {NoStop}%
\bibitem [{\citenamefont {Abbas}\ \emph {et~al.}(2021)\citenamefont {Abbas}, \citenamefont {Sutter}, \citenamefont {Zoufal}, \citenamefont {Lucchi}, \citenamefont {Figalli},\ and\ \citenamefont {Woerner}}]{NNs_Abbas2021}%
  \BibitemOpen
  \bibfield  {author} {\bibinfo {author} {\bibfnamefont {A.}~\bibnamefont {Abbas}}, \bibinfo {author} {\bibfnamefont {D.}~\bibnamefont {Sutter}}, \bibinfo {author} {\bibfnamefont {C.}~\bibnamefont {Zoufal}}, \bibinfo {author} {\bibfnamefont {A.}~\bibnamefont {Lucchi}}, \bibinfo {author} {\bibfnamefont {A.}~\bibnamefont {Figalli}},\ and\ \bibinfo {author} {\bibfnamefont {S.}~\bibnamefont {Woerner}},\ }\bibfield  {title} {\bibinfo {title} {The power of quantum neural networks},\ }\href {https://doi.org/10.1038/s43588-021-00084-1} {\bibfield  {journal} {\bibinfo  {journal} {Nature Computational Science}\ }\textbf {\bibinfo {volume} {1}},\ \bibinfo {pages} {403} (\bibinfo {year} {2021})}\BibitemShut {NoStop}%
\bibitem [{\citenamefont {Cerezo}\ \emph {et~al.}(2021)\citenamefont {Cerezo}, \citenamefont {Arrasmith}, \citenamefont {Babbush}, \citenamefont {Benjamin}, \citenamefont {Endo}, \citenamefont {Fujii}, \citenamefont {McClean}, \citenamefont {Mitarai}, \citenamefont {Yuan}, \citenamefont {Cincio},\ and\ \citenamefont {Coles}}]{CerezoVQE}%
  \BibitemOpen
  \bibfield  {author} {\bibinfo {author} {\bibfnamefont {M.}~\bibnamefont {Cerezo}}, \bibinfo {author} {\bibfnamefont {A.}~\bibnamefont {Arrasmith}}, \bibinfo {author} {\bibfnamefont {R.}~\bibnamefont {Babbush}}, \bibinfo {author} {\bibfnamefont {S.~C.}\ \bibnamefont {Benjamin}}, \bibinfo {author} {\bibfnamefont {S.}~\bibnamefont {Endo}}, \bibinfo {author} {\bibfnamefont {K.}~\bibnamefont {Fujii}}, \bibinfo {author} {\bibfnamefont {J.~R.}\ \bibnamefont {McClean}}, \bibinfo {author} {\bibfnamefont {K.}~\bibnamefont {Mitarai}}, \bibinfo {author} {\bibfnamefont {X.}~\bibnamefont {Yuan}}, \bibinfo {author} {\bibfnamefont {L.}~\bibnamefont {Cincio}},\ and\ \bibinfo {author} {\bibfnamefont {P.~J.}\ \bibnamefont {Coles}},\ }\bibfield  {title} {\bibinfo {title} {Variational quantum algorithms},\ }\href {https://doi.org/10.1038/s42254-021-00348-9} {\bibfield  {journal} {\bibinfo  {journal} {Nature Reviews Physics}\ }\textbf {\bibinfo {volume} {3}},\ \bibinfo {pages} {625} (\bibinfo {year} {2021})}\BibitemShut
  {NoStop}%
\bibitem [{\citenamefont {Reck}\ \emph {et~al.}(1994)\citenamefont {Reck}, \citenamefont {Zeilinger}, \citenamefont {Bernstein},\ and\ \citenamefont {Bertani}}]{ReckDesign}%
  \BibitemOpen
  \bibfield  {author} {\bibinfo {author} {\bibfnamefont {M.}~\bibnamefont {Reck}}, \bibinfo {author} {\bibfnamefont {A.}~\bibnamefont {Zeilinger}}, \bibinfo {author} {\bibfnamefont {H.~J.}\ \bibnamefont {Bernstein}},\ and\ \bibinfo {author} {\bibfnamefont {P.}~\bibnamefont {Bertani}},\ }\bibfield  {title} {\bibinfo {title} {Experimental realization of any discrete unitary operator},\ }\href {https://doi.org/10.1103/PhysRevLett.73.58} {\bibfield  {journal} {\bibinfo  {journal} {Phys. Rev. Lett.}\ }\textbf {\bibinfo {volume} {73}},\ \bibinfo {pages} {58} (\bibinfo {year} {1994})}\BibitemShut {NoStop}%
\bibitem [{\citenamefont {Clements}\ \emph {et~al.}(2016)\citenamefont {Clements}, \citenamefont {Humphreys}, \citenamefont {Metcalf}, \citenamefont {Kolthammer},\ and\ \citenamefont {Walmsley}}]{ClementsDesign}%
  \BibitemOpen
  \bibfield  {author} {\bibinfo {author} {\bibfnamefont {W.~R.}\ \bibnamefont {Clements}}, \bibinfo {author} {\bibfnamefont {P.~C.}\ \bibnamefont {Humphreys}}, \bibinfo {author} {\bibfnamefont {B.~J.}\ \bibnamefont {Metcalf}}, \bibinfo {author} {\bibfnamefont {W.~S.}\ \bibnamefont {Kolthammer}},\ and\ \bibinfo {author} {\bibfnamefont {I.~A.}\ \bibnamefont {Walmsley}},\ }\bibfield  {title} {\bibinfo {title} {Optimal design for universal multiport interferometers},\ }\href {https://doi.org/10.1364/OPTICA.3.001460} {\bibfield  {journal} {\bibinfo  {journal} {Optica}\ }\textbf {\bibinfo {volume} {3}},\ \bibinfo {pages} {1460} (\bibinfo {year} {2016})}\BibitemShut {NoStop}%
\bibitem [{\citenamefont {Motes}\ \emph {et~al.}(2014)\citenamefont {Motes}, \citenamefont {Gilchrist}, \citenamefont {Dowling},\ and\ \citenamefont {Rohde}}]{TimeBin_interferometer}%
  \BibitemOpen
  \bibfield  {author} {\bibinfo {author} {\bibfnamefont {K.~R.}\ \bibnamefont {Motes}}, \bibinfo {author} {\bibfnamefont {A.}~\bibnamefont {Gilchrist}}, \bibinfo {author} {\bibfnamefont {J.~P.}\ \bibnamefont {Dowling}},\ and\ \bibinfo {author} {\bibfnamefont {P.~P.}\ \bibnamefont {Rohde}},\ }\bibfield  {title} {\bibinfo {title} {Scalable boson sampling with time-bin encoding using a loop-based architecture},\ }\href {https://doi.org/10.1103/PhysRevLett.113.120501} {\bibfield  {journal} {\bibinfo  {journal} {Phys. Rev. Lett.}\ }\textbf {\bibinfo {volume} {113}},\ \bibinfo {pages} {120501} (\bibinfo {year} {2014})}\BibitemShut {NoStop}%
\bibitem [{\citenamefont {Saygin}\ \emph {et~al.}(2020)\citenamefont {Saygin}, \citenamefont {Kondratyev}, \citenamefont {Dyakonov}, \citenamefont {Mironov}, \citenamefont {Straupe},\ and\ \citenamefont {Kulik}}]{RobustSaygin}%
  \BibitemOpen
  \bibfield  {author} {\bibinfo {author} {\bibfnamefont {M.~Y.}\ \bibnamefont {Saygin}}, \bibinfo {author} {\bibfnamefont {I.~V.}\ \bibnamefont {Kondratyev}}, \bibinfo {author} {\bibfnamefont {I.~V.}\ \bibnamefont {Dyakonov}}, \bibinfo {author} {\bibfnamefont {S.~A.}\ \bibnamefont {Mironov}}, \bibinfo {author} {\bibfnamefont {S.~S.}\ \bibnamefont {Straupe}},\ and\ \bibinfo {author} {\bibfnamefont {S.~P.}\ \bibnamefont {Kulik}},\ }\bibfield  {title} {\bibinfo {title} {Robust architecture for programmable universal unitaries},\ }\href {https://doi.org/10.1103/PhysRevLett.124.010501} {\bibfield  {journal} {\bibinfo  {journal} {Phys. Rev. Lett.}\ }\textbf {\bibinfo {volume} {124}},\ \bibinfo {pages} {010501} (\bibinfo {year} {2020})}\BibitemShut {NoStop}%
\bibitem [{\citenamefont {Fldzhyan}\ \emph {et~al.}(2020)\citenamefont {Fldzhyan}, \citenamefont {Saygin},\ and\ \citenamefont {Kulik}}]{FldzhyanDesign}%
  \BibitemOpen
  \bibfield  {author} {\bibinfo {author} {\bibfnamefont {S.~A.}\ \bibnamefont {Fldzhyan}}, \bibinfo {author} {\bibfnamefont {M.~Y.}\ \bibnamefont {Saygin}},\ and\ \bibinfo {author} {\bibfnamefont {S.~P.}\ \bibnamefont {Kulik}},\ }\bibfield  {title} {\bibinfo {title} {Optimal design of error-tolerant reprogrammable multiport interferometers},\ }\href {https://doi.org/10.1364/OL.385433} {\bibfield  {journal} {\bibinfo  {journal} {Opt. Lett.}\ }\textbf {\bibinfo {volume} {45}},\ \bibinfo {pages} {2632} (\bibinfo {year} {2020})}\BibitemShut {NoStop}%
\bibitem [{\citenamefont {Taballione}\ \emph {et~al.}(2023)\citenamefont {Taballione}, \citenamefont {Anguita}, \citenamefont {de~Goede}, \citenamefont {Venderbosch}, \citenamefont {Kassenberg}, \citenamefont {Snijders}, \citenamefont {Kannan}, \citenamefont {Vleeshouwers}, \citenamefont {Smith}, \citenamefont {Epping}, \citenamefont {van~der Meer}, \citenamefont {Pinkse}, \citenamefont {van~den Vlekkert},\ and\ \citenamefont {Renema}}]{Taballione2023modeuniversal}%
  \BibitemOpen
  \bibfield  {author} {\bibinfo {author} {\bibfnamefont {C.}~\bibnamefont {Taballione}}, \bibinfo {author} {\bibfnamefont {M.~C.}\ \bibnamefont {Anguita}}, \bibinfo {author} {\bibfnamefont {M.}~\bibnamefont {de~Goede}}, \bibinfo {author} {\bibfnamefont {P.}~\bibnamefont {Venderbosch}}, \bibinfo {author} {\bibfnamefont {B.}~\bibnamefont {Kassenberg}}, \bibinfo {author} {\bibfnamefont {H.}~\bibnamefont {Snijders}}, \bibinfo {author} {\bibfnamefont {N.}~\bibnamefont {Kannan}}, \bibinfo {author} {\bibfnamefont {W.~L.}\ \bibnamefont {Vleeshouwers}}, \bibinfo {author} {\bibfnamefont {D.}~\bibnamefont {Smith}}, \bibinfo {author} {\bibfnamefont {J.~P.}\ \bibnamefont {Epping}}, \bibinfo {author} {\bibfnamefont {R.}~\bibnamefont {van~der Meer}}, \bibinfo {author} {\bibfnamefont {P.~W.~H.}\ \bibnamefont {Pinkse}}, \bibinfo {author} {\bibfnamefont {H.}~\bibnamefont {van~den Vlekkert}},\ and\ \bibinfo {author} {\bibfnamefont {J.~J.}\ \bibnamefont {Renema}},\ }\bibfield  {title} {\bibinfo {title} {20-{M}ode {U}niversal
  {Q}uantum {P}hotonic {P}rocessor},\ }\href {https://doi.org/10.22331/q-2023-08-01-1071} {\bibfield  {journal} {\bibinfo  {journal} {{Quantum}}\ }\textbf {\bibinfo {volume} {7}},\ \bibinfo {pages} {1071} (\bibinfo {year} {2023})}\BibitemShut {NoStop}%
\bibitem [{\citenamefont {Kondratyev}\ \emph {et~al.}(2023)\citenamefont {Kondratyev}, \citenamefont {Ivanova}, \citenamefont {Zhuravitskii}, \citenamefont {Argenchiev}, \citenamefont {Skryabin}, \citenamefont {Dyakonov}, \citenamefont {Fldzhyan}, \citenamefont {Saygin}, \citenamefont {Straupe}, \citenamefont {Korneev},\ and\ \citenamefont {Kulik}}]{KondratyevLargeScale}%
  \BibitemOpen
  \bibfield  {author} {\bibinfo {author} {\bibfnamefont {I.~V.}\ \bibnamefont {Kondratyev}}, \bibinfo {author} {\bibfnamefont {V.~V.}\ \bibnamefont {Ivanova}}, \bibinfo {author} {\bibfnamefont {S.~A.}\ \bibnamefont {Zhuravitskii}}, \bibinfo {author} {\bibfnamefont {A.~S.}\ \bibnamefont {Argenchiev}}, \bibinfo {author} {\bibfnamefont {N.~N.}\ \bibnamefont {Skryabin}}, \bibinfo {author} {\bibfnamefont {I.~V.}\ \bibnamefont {Dyakonov}}, \bibinfo {author} {\bibfnamefont {S.~A.}\ \bibnamefont {Fldzhyan}}, \bibinfo {author} {\bibfnamefont {M.~Y.}\ \bibnamefont {Saygin}}, \bibinfo {author} {\bibfnamefont {S.~S.}\ \bibnamefont {Straupe}}, \bibinfo {author} {\bibfnamefont {A.~A.}\ \bibnamefont {Korneev}},\ and\ \bibinfo {author} {\bibfnamefont {S.~P.}\ \bibnamefont {Kulik}},\ }\href@noop {} {\bibinfo {title} {Large-scale error-tolerant programmable interferometer fabricated by femtosecond laser writing}} (\bibinfo {year} {2023}),\ \Eprint {https://arxiv.org/abs/2308.13452} {arXiv:2308.13452 [quant-ph]} \BibitemShut
  {NoStop}%
\bibitem [{\citenamefont {Wang}\ \emph {et~al.}(2019)\citenamefont {Wang}, \citenamefont {Qin}, \citenamefont {Ding}, \citenamefont {Chen}, \citenamefont {Chen}, \citenamefont {You}, \citenamefont {He}, \citenamefont {Jiang}, \citenamefont {You}, \citenamefont {Wang}, \citenamefont {Schneider}, \citenamefont {Renema}, \citenamefont {H\"ofling}, \citenamefont {Lu},\ and\ \citenamefont {Pan}}]{BOSON_SAMPLING_DV}%
  \BibitemOpen
  \bibfield  {author} {\bibinfo {author} {\bibfnamefont {H.}~\bibnamefont {Wang}}, \bibinfo {author} {\bibfnamefont {J.}~\bibnamefont {Qin}}, \bibinfo {author} {\bibfnamefont {X.}~\bibnamefont {Ding}}, \bibinfo {author} {\bibfnamefont {M.-C.}\ \bibnamefont {Chen}}, \bibinfo {author} {\bibfnamefont {S.}~\bibnamefont {Chen}}, \bibinfo {author} {\bibfnamefont {X.}~\bibnamefont {You}}, \bibinfo {author} {\bibfnamefont {Y.-M.}\ \bibnamefont {He}}, \bibinfo {author} {\bibfnamefont {X.}~\bibnamefont {Jiang}}, \bibinfo {author} {\bibfnamefont {L.}~\bibnamefont {You}}, \bibinfo {author} {\bibfnamefont {Z.}~\bibnamefont {Wang}}, \bibinfo {author} {\bibfnamefont {C.}~\bibnamefont {Schneider}}, \bibinfo {author} {\bibfnamefont {J.~J.}\ \bibnamefont {Renema}}, \bibinfo {author} {\bibfnamefont {S.}~\bibnamefont {H\"ofling}}, \bibinfo {author} {\bibfnamefont {C.-Y.}\ \bibnamefont {Lu}},\ and\ \bibinfo {author} {\bibfnamefont {J.-W.}\ \bibnamefont {Pan}},\ }\bibfield  {title} {\bibinfo {title} {Boson sampling with 20 input
  photons and a 60-mode interferometer in a $1{0}^{14}$-dimensional hilbert space},\ }\href {https://doi.org/10.1103/PhysRevLett.123.250503} {\bibfield  {journal} {\bibinfo  {journal} {Phys. Rev. Lett.}\ }\textbf {\bibinfo {volume} {123}},\ \bibinfo {pages} {250503} (\bibinfo {year} {2019})}\BibitemShut {NoStop}%
\bibitem [{\citenamefont {Knill}\ \emph {et~al.}(2001)\citenamefont {Knill}, \citenamefont {Laflamme},\ and\ \citenamefont {Milburn}}]{KLM}%
  \BibitemOpen
  \bibfield  {author} {\bibinfo {author} {\bibfnamefont {E.}~\bibnamefont {Knill}}, \bibinfo {author} {\bibfnamefont {R.}~\bibnamefont {Laflamme}},\ and\ \bibinfo {author} {\bibfnamefont {G.~J.}\ \bibnamefont {Milburn}},\ }\bibfield  {title} {\bibinfo {title} {A scheme for efficient quantum computation with linear optics},\ }\href {https://doi.org/10.1038/35051009} {\bibfield  {journal} {\bibinfo  {journal} {Nature}\ }\textbf {\bibinfo {volume} {409}},\ \bibinfo {pages} {46} (\bibinfo {year} {2001})}\BibitemShut {NoStop}%
\bibitem [{\citenamefont {Briegel}\ \emph {et~al.}(2009)\citenamefont {Briegel}, \citenamefont {Browne}, \citenamefont {D{\"u}r}, \citenamefont {Raussendorf},\ and\ \citenamefont {Van~den Nest}}]{MBQC}%
  \BibitemOpen
  \bibfield  {author} {\bibinfo {author} {\bibfnamefont {H.~J.}\ \bibnamefont {Briegel}}, \bibinfo {author} {\bibfnamefont {D.~E.}\ \bibnamefont {Browne}}, \bibinfo {author} {\bibfnamefont {W.}~\bibnamefont {D{\"u}r}}, \bibinfo {author} {\bibfnamefont {R.}~\bibnamefont {Raussendorf}},\ and\ \bibinfo {author} {\bibfnamefont {M.}~\bibnamefont {Van~den Nest}},\ }\bibfield  {title} {\bibinfo {title} {Measurement-based quantum computation},\ }\href {https://doi.org/10.1038/nphys1157} {\bibfield  {journal} {\bibinfo  {journal} {Nature Physics}\ }\textbf {\bibinfo {volume} {5}},\ \bibinfo {pages} {19} (\bibinfo {year} {2009})}\BibitemShut {NoStop}%
\bibitem [{\citenamefont {Bartolucci}\ \emph {et~al.}(2023)\citenamefont {Bartolucci}, \citenamefont {Birchall}, \citenamefont {Bomb{\'i}n}, \citenamefont {Cable}, \citenamefont {Dawson}, \citenamefont {Gimeno-Segovia}, \citenamefont {Johnston}, \citenamefont {Kieling}, \citenamefont {Nickerson}, \citenamefont {Pant}, \citenamefont {Pastawski}, \citenamefont {Rudolph},\ and\ \citenamefont {Sparrow}}]{FBQC}%
  \BibitemOpen
  \bibfield  {author} {\bibinfo {author} {\bibfnamefont {S.}~\bibnamefont {Bartolucci}}, \bibinfo {author} {\bibfnamefont {P.}~\bibnamefont {Birchall}}, \bibinfo {author} {\bibfnamefont {H.}~\bibnamefont {Bomb{\'i}n}}, \bibinfo {author} {\bibfnamefont {H.}~\bibnamefont {Cable}}, \bibinfo {author} {\bibfnamefont {C.}~\bibnamefont {Dawson}}, \bibinfo {author} {\bibfnamefont {M.}~\bibnamefont {Gimeno-Segovia}}, \bibinfo {author} {\bibfnamefont {E.}~\bibnamefont {Johnston}}, \bibinfo {author} {\bibfnamefont {K.}~\bibnamefont {Kieling}}, \bibinfo {author} {\bibfnamefont {N.}~\bibnamefont {Nickerson}}, \bibinfo {author} {\bibfnamefont {M.}~\bibnamefont {Pant}}, \bibinfo {author} {\bibfnamefont {F.}~\bibnamefont {Pastawski}}, \bibinfo {author} {\bibfnamefont {T.}~\bibnamefont {Rudolph}},\ and\ \bibinfo {author} {\bibfnamefont {C.}~\bibnamefont {Sparrow}},\ }\bibfield  {title} {\bibinfo {title} {Fusion-based quantum computation},\ }\href {https://doi.org/10.1038/s41467-023-36493-1} {\bibfield  {journal} {\bibinfo
  {journal} {Nature Communications}\ }\textbf {\bibinfo {volume} {14}},\ \bibinfo {pages} {912} (\bibinfo {year} {2023})}\BibitemShut {NoStop}%
\bibitem [{\citenamefont {Volz}\ \emph {et~al.}(2014)\citenamefont {Volz}, \citenamefont {Scheucher}, \citenamefont {Junge},\ and\ \citenamefont {Rauschenbeutel}}]{RB_NPS_WISPERRING}%
  \BibitemOpen
  \bibfield  {author} {\bibinfo {author} {\bibfnamefont {J.}~\bibnamefont {Volz}}, \bibinfo {author} {\bibfnamefont {M.}~\bibnamefont {Scheucher}}, \bibinfo {author} {\bibfnamefont {C.}~\bibnamefont {Junge}},\ and\ \bibinfo {author} {\bibfnamefont {A.}~\bibnamefont {Rauschenbeutel}},\ }\bibfield  {title} {\bibinfo {title} {Nonlinear $\pi$ phase shift for single fibre-guided photons interacting with a single resonator-enhanced atom},\ }\href {https://doi.org/10.1038/nphoton.2014.253} {\bibfield  {journal} {\bibinfo  {journal} {Nature Photonics}\ }\textbf {\bibinfo {volume} {8}},\ \bibinfo {pages} {965} (\bibinfo {year} {2014})}\BibitemShut {NoStop}%
\bibitem [{\citenamefont {Feizpour}\ \emph {et~al.}(2015)\citenamefont {Feizpour}, \citenamefont {Hallaji}, \citenamefont {Dmochowski},\ and\ \citenamefont {Steinberg}}]{Vapour_NS}%
  \BibitemOpen
  \bibfield  {author} {\bibinfo {author} {\bibfnamefont {A.}~\bibnamefont {Feizpour}}, \bibinfo {author} {\bibfnamefont {M.}~\bibnamefont {Hallaji}}, \bibinfo {author} {\bibfnamefont {G.}~\bibnamefont {Dmochowski}},\ and\ \bibinfo {author} {\bibfnamefont {A.~M.}\ \bibnamefont {Steinberg}},\ }\bibfield  {title} {\bibinfo {title} {Observation of the nonlinear phase shift due to single post-selected photons},\ }\href {https://doi.org/10.1038/nphys3433} {\bibfield  {journal} {\bibinfo  {journal} {Nature Physics}\ }\textbf {\bibinfo {volume} {11}},\ \bibinfo {pages} {905} (\bibinfo {year} {2015})}\BibitemShut {NoStop}%
\bibitem [{\citenamefont {Javadi}\ \emph {et~al.}(2015)\citenamefont {Javadi}, \citenamefont {S{\"o}llner}, \citenamefont {Arcari}, \citenamefont {Hansen}, \citenamefont {Midolo}, \citenamefont {Mahmoodian}, \citenamefont {Kir{\v{s}}ansk{\.{e}}}, \citenamefont {Pregnolato}, \citenamefont {Lee}, \citenamefont {Song}, \citenamefont {Stobbe},\ and\ \citenamefont {Lodahl}}]{SinglePhoton_NL_nature}%
  \BibitemOpen
  \bibfield  {author} {\bibinfo {author} {\bibfnamefont {A.}~\bibnamefont {Javadi}}, \bibinfo {author} {\bibfnamefont {I.}~\bibnamefont {S{\"o}llner}}, \bibinfo {author} {\bibfnamefont {M.}~\bibnamefont {Arcari}}, \bibinfo {author} {\bibfnamefont {S.~L.}\ \bibnamefont {Hansen}}, \bibinfo {author} {\bibfnamefont {L.}~\bibnamefont {Midolo}}, \bibinfo {author} {\bibfnamefont {S.}~\bibnamefont {Mahmoodian}}, \bibinfo {author} {\bibfnamefont {G.}~\bibnamefont {Kir{\v{s}}ansk{\.{e}}}}, \bibinfo {author} {\bibfnamefont {T.}~\bibnamefont {Pregnolato}}, \bibinfo {author} {\bibfnamefont {E.~H.}\ \bibnamefont {Lee}}, \bibinfo {author} {\bibfnamefont {J.~D.}\ \bibnamefont {Song}}, \bibinfo {author} {\bibfnamefont {S.}~\bibnamefont {Stobbe}},\ and\ \bibinfo {author} {\bibfnamefont {P.}~\bibnamefont {Lodahl}},\ }\bibfield  {title} {\bibinfo {title} {Single-photon non-linear optics with a quantum dot in a waveguide},\ }\href {https://doi.org/10.1038/ncomms9655} {\bibfield  {journal} {\bibinfo  {journal} {Nature
  Communications}\ }\textbf {\bibinfo {volume} {6}},\ \bibinfo {pages} {8655} (\bibinfo {year} {2015})}\BibitemShut {NoStop}%
\bibitem [{\citenamefont {Staunstrup}\ \emph {et~al.}(2023)\citenamefont {Staunstrup}, \citenamefont {Tiranov}, \citenamefont {Wang}, \citenamefont {Scholz}, \citenamefont {Wieck}, \citenamefont {Ludwig}, \citenamefont {Midolo}, \citenamefont {Rotenberg}, \citenamefont {Lodahl},\ and\ \citenamefont {Jeannic}}]{Lodahl_arxiv_NS}%
  \BibitemOpen
  \bibfield  {author} {\bibinfo {author} {\bibfnamefont {M.~J.~R.}\ \bibnamefont {Staunstrup}}, \bibinfo {author} {\bibfnamefont {A.}~\bibnamefont {Tiranov}}, \bibinfo {author} {\bibfnamefont {Y.}~\bibnamefont {Wang}}, \bibinfo {author} {\bibfnamefont {S.}~\bibnamefont {Scholz}}, \bibinfo {author} {\bibfnamefont {A.~D.}\ \bibnamefont {Wieck}}, \bibinfo {author} {\bibfnamefont {A.}~\bibnamefont {Ludwig}}, \bibinfo {author} {\bibfnamefont {L.}~\bibnamefont {Midolo}}, \bibinfo {author} {\bibfnamefont {N.}~\bibnamefont {Rotenberg}}, \bibinfo {author} {\bibfnamefont {P.}~\bibnamefont {Lodahl}},\ and\ \bibinfo {author} {\bibfnamefont {H.~L.}\ \bibnamefont {Jeannic}},\ }\href@noop {} {\bibinfo {title} {Direct observation of non-linear optical phase shift induced by a single quantum emitter in a waveguide}} (\bibinfo {year} {2023}),\ \Eprint {https://arxiv.org/abs/2305.06839} {arXiv:2305.06839 [quant-ph]} \BibitemShut {NoStop}%
\bibitem [{\citenamefont {Zhao}\ and\ \citenamefont {Fang}(2022)}]{Bulk_nonlinearity}%
  \BibitemOpen
  \bibfield  {author} {\bibinfo {author} {\bibfnamefont {M.}~\bibnamefont {Zhao}}\ and\ \bibinfo {author} {\bibfnamefont {K.}~\bibnamefont {Fang}},\ }\bibfield  {title} {\bibinfo {title} {Ingap quantum nanophotonic integrated circuits with 1.5\% nonlinearity-to-loss ratio},\ }\href {https://doi.org/10.1364/OPTICA.440383} {\bibfield  {journal} {\bibinfo  {journal} {Optica}\ }\textbf {\bibinfo {volume} {9}},\ \bibinfo {pages} {258} (\bibinfo {year} {2022})}\BibitemShut {NoStop}%
\bibitem [{\citenamefont {Yanagimoto}\ \emph {et~al.}(2022)\citenamefont {Yanagimoto}, \citenamefont {Ng}, \citenamefont {Jankowski}, \citenamefont {Mabuchi},\ and\ \citenamefont {Hamerly}}]{TemporalTrapping}%
  \BibitemOpen
  \bibfield  {author} {\bibinfo {author} {\bibfnamefont {R.}~\bibnamefont {Yanagimoto}}, \bibinfo {author} {\bibfnamefont {E.}~\bibnamefont {Ng}}, \bibinfo {author} {\bibfnamefont {M.}~\bibnamefont {Jankowski}}, \bibinfo {author} {\bibfnamefont {H.}~\bibnamefont {Mabuchi}},\ and\ \bibinfo {author} {\bibfnamefont {R.}~\bibnamefont {Hamerly}},\ }\bibfield  {title} {\bibinfo {title} {Temporal trapping: a route to strong coupling and deterministic optical quantum computation},\ }\href {https://doi.org/10.1364/OPTICA.473276} {\bibfield  {journal} {\bibinfo  {journal} {Optica}\ }\textbf {\bibinfo {volume} {9}},\ \bibinfo {pages} {1289} (\bibinfo {year} {2022})}\BibitemShut {NoStop}%
\bibitem [{\citenamefont {O'Brien}\ \emph {et~al.}(2009)\citenamefont {O'Brien}, \citenamefont {Furusawa},\ and\ \citenamefont {Vu{\v{c}}kovi{\'{c}}}}]{PhotonicQuantumTechnologies}%
  \BibitemOpen
  \bibfield  {author} {\bibinfo {author} {\bibfnamefont {J.~L.}\ \bibnamefont {O'Brien}}, \bibinfo {author} {\bibfnamefont {A.}~\bibnamefont {Furusawa}},\ and\ \bibinfo {author} {\bibfnamefont {J.}~\bibnamefont {Vu{\v{c}}kovi{\'{c}}}},\ }\bibfield  {title} {\bibinfo {title} {Photonic quantum technologies},\ }\href {https://doi.org/10.1038/nphoton.2009.229} {\bibfield  {journal} {\bibinfo  {journal} {Nature Photonics}\ }\textbf {\bibinfo {volume} {3}},\ \bibinfo {pages} {687} (\bibinfo {year} {2009})}\BibitemShut {NoStop}%
\bibitem [{\citenamefont {Adhikari}\ \emph {et~al.}(2013)\citenamefont {Adhikari}, \citenamefont {Hafezi},\ and\ \citenamefont {Taylor}}]{Superconducting_NL}%
  \BibitemOpen
  \bibfield  {author} {\bibinfo {author} {\bibfnamefont {P.}~\bibnamefont {Adhikari}}, \bibinfo {author} {\bibfnamefont {M.}~\bibnamefont {Hafezi}},\ and\ \bibinfo {author} {\bibfnamefont {J.~M.}\ \bibnamefont {Taylor}},\ }\bibfield  {title} {\bibinfo {title} {Nonlinear optics quantum computing with circuit qed},\ }\href {https://doi.org/10.1103/PhysRevLett.110.060503} {\bibfield  {journal} {\bibinfo  {journal} {Phys. Rev. Lett.}\ }\textbf {\bibinfo {volume} {110}},\ \bibinfo {pages} {060503} (\bibinfo {year} {2013})}\BibitemShut {NoStop}%
\bibitem [{\citenamefont {LeCun}\ \emph {et~al.}(2015)\citenamefont {LeCun}, \citenamefont {Bengio},\ and\ \citenamefont {Hinton}}]{DeepLearning}%
  \BibitemOpen
  \bibfield  {author} {\bibinfo {author} {\bibfnamefont {Y.}~\bibnamefont {LeCun}}, \bibinfo {author} {\bibfnamefont {Y.}~\bibnamefont {Bengio}},\ and\ \bibinfo {author} {\bibfnamefont {G.}~\bibnamefont {Hinton}},\ }\bibfield  {title} {\bibinfo {title} {Deep learning},\ }\href {https://doi.org/10.1038/nature14539} {\bibfield  {journal} {\bibinfo  {journal} {Nature}\ }\textbf {\bibinfo {volume} {521}},\ \bibinfo {pages} {436} (\bibinfo {year} {2015})}\BibitemShut {NoStop}%
\bibitem [{\citenamefont {Steinbrecher}\ \emph {et~al.}(2019)\citenamefont {Steinbrecher}, \citenamefont {Olson}, \citenamefont {Englund},\ and\ \citenamefont {Carolan}}]{QuantumOpticalNNs}%
  \BibitemOpen
  \bibfield  {author} {\bibinfo {author} {\bibfnamefont {G.~R.}\ \bibnamefont {Steinbrecher}}, \bibinfo {author} {\bibfnamefont {J.~P.}\ \bibnamefont {Olson}}, \bibinfo {author} {\bibfnamefont {D.}~\bibnamefont {Englund}},\ and\ \bibinfo {author} {\bibfnamefont {J.}~\bibnamefont {Carolan}},\ }\bibfield  {title} {\bibinfo {title} {Quantum optical neural networks},\ }\href {https://doi.org/10.1038/s41534-019-0174-7} {\bibfield  {journal} {\bibinfo  {journal} {npj Quantum Information}\ }\textbf {\bibinfo {volume} {5}},\ \bibinfo {pages} {60} (\bibinfo {year} {2019})}\BibitemShut {NoStop}%
\bibitem [{\citenamefont {Ewaniuk}\ \emph {et~al.}(2023)\citenamefont {Ewaniuk}, \citenamefont {Carolan}, \citenamefont {Shastri},\ and\ \citenamefont {Rotenberg}}]{EwaniukQONNs}%
  \BibitemOpen
  \bibfield  {author} {\bibinfo {author} {\bibfnamefont {J.}~\bibnamefont {Ewaniuk}}, \bibinfo {author} {\bibfnamefont {J.}~\bibnamefont {Carolan}}, \bibinfo {author} {\bibfnamefont {B.~J.}\ \bibnamefont {Shastri}},\ and\ \bibinfo {author} {\bibfnamefont {N.}~\bibnamefont {Rotenberg}},\ }\bibfield  {title} {\bibinfo {title} {Imperfect quantum photonic neural networks},\ }\href {https://doi.org/https://doi.org/10.1002/qute.202200125} {\bibfield  {journal} {\bibinfo  {journal} {Advanced Quantum Technologies}\ }\textbf {\bibinfo {volume} {6}},\ \bibinfo {pages} {2200125} (\bibinfo {year} {2023})}\BibitemShut {NoStop}%
\bibitem [{\citenamefont {Slussarenko}\ and\ \citenamefont {Pryde}(2019)}]{Slussarenko}%
  \BibitemOpen
  \bibfield  {author} {\bibinfo {author} {\bibfnamefont {S.}~\bibnamefont {Slussarenko}}\ and\ \bibinfo {author} {\bibfnamefont {G.~J.}\ \bibnamefont {Pryde}},\ }\bibfield  {title} {\bibinfo {title} {{Photonic quantum information processing: A concise review}},\ }\href {https://doi.org/10.1063/1.5115814} {\bibfield  {journal} {\bibinfo  {journal} {Applied Physics Reviews}\ }\textbf {\bibinfo {volume} {6}},\ \bibinfo {pages} {041303} (\bibinfo {year} {2019})},\ \Eprint {https://arxiv.org/abs/https://pubs.aip.org/aip/apr/article-pdf/doi/10.1063/1.5115814/19739502/041303\_1\_online.pdf} {https://pubs.aip.org/aip/apr/article-pdf/doi/10.1063/1.5115814/19739502/041303\_1\_online.pdf} \BibitemShut {NoStop}%
\bibitem [{\citenamefont {Wang}\ \emph {et~al.}(2020)\citenamefont {Wang}, \citenamefont {Sciarrino}, \citenamefont {Laing},\ and\ \citenamefont {Thompson}}]{IntegratedPICreview}%
  \BibitemOpen
  \bibfield  {author} {\bibinfo {author} {\bibfnamefont {J.}~\bibnamefont {Wang}}, \bibinfo {author} {\bibfnamefont {F.}~\bibnamefont {Sciarrino}}, \bibinfo {author} {\bibfnamefont {A.}~\bibnamefont {Laing}},\ and\ \bibinfo {author} {\bibfnamefont {M.~G.}\ \bibnamefont {Thompson}},\ }\bibfield  {title} {\bibinfo {title} {Integrated photonic quantum technologies},\ }\href {https://doi.org/10.1038/s41566-019-0532-1} {\bibfield  {journal} {\bibinfo  {journal} {Nature Photonics}\ }\textbf {\bibinfo {volume} {14}},\ \bibinfo {pages} {273} (\bibinfo {year} {2020})}\BibitemShut {NoStop}%
\bibitem [{\citenamefont {Heindel}\ \emph {et~al.}(2023)\citenamefont {Heindel}, \citenamefont {Kim}, \citenamefont {Gregersen}, \citenamefont {Rastelli},\ and\ \citenamefont {Reitzenstein}}]{QDforQIP}%
  \BibitemOpen
  \bibfield  {author} {\bibinfo {author} {\bibfnamefont {T.}~\bibnamefont {Heindel}}, \bibinfo {author} {\bibfnamefont {J.-H.}\ \bibnamefont {Kim}}, \bibinfo {author} {\bibfnamefont {N.}~\bibnamefont {Gregersen}}, \bibinfo {author} {\bibfnamefont {A.}~\bibnamefont {Rastelli}},\ and\ \bibinfo {author} {\bibfnamefont {S.}~\bibnamefont {Reitzenstein}},\ }\bibfield  {title} {\bibinfo {title} {Quantum dots for photonic quantum information technology},\ }\href {https://doi.org/10.1364/AOP.490091} {\bibfield  {journal} {\bibinfo  {journal} {Adv. Opt. Photon.}\ }\textbf {\bibinfo {volume} {15}},\ \bibinfo {pages} {613} (\bibinfo {year} {2023})}\BibitemShut {NoStop}%
\bibitem [{\citenamefont {Wang}\ \emph {et~al.}(2023)\citenamefont {Wang}, \citenamefont {Faurby}, \citenamefont {Ruf}, \citenamefont {Sund}, \citenamefont {Nielsen}, \citenamefont {Volet}, \citenamefont {Heck}, \citenamefont {Bart}, \citenamefont {Wieck}, \citenamefont {Ludwig}, \citenamefont {Midolo}, \citenamefont {Paesani},\ and\ \citenamefont {Lodahl}}]{LodahlSource}%
  \BibitemOpen
  \bibfield  {author} {\bibinfo {author} {\bibfnamefont {Y.}~\bibnamefont {Wang}}, \bibinfo {author} {\bibfnamefont {C.~F.~D.}\ \bibnamefont {Faurby}}, \bibinfo {author} {\bibfnamefont {F.}~\bibnamefont {Ruf}}, \bibinfo {author} {\bibfnamefont {P.~I.}\ \bibnamefont {Sund}}, \bibinfo {author} {\bibfnamefont {K.}~\bibnamefont {Nielsen}}, \bibinfo {author} {\bibfnamefont {N.}~\bibnamefont {Volet}}, \bibinfo {author} {\bibfnamefont {M.~J.~R.}\ \bibnamefont {Heck}}, \bibinfo {author} {\bibfnamefont {N.}~\bibnamefont {Bart}}, \bibinfo {author} {\bibfnamefont {A.~D.}\ \bibnamefont {Wieck}}, \bibinfo {author} {\bibfnamefont {A.}~\bibnamefont {Ludwig}}, \bibinfo {author} {\bibfnamefont {L.}~\bibnamefont {Midolo}}, \bibinfo {author} {\bibfnamefont {S.}~\bibnamefont {Paesani}},\ and\ \bibinfo {author} {\bibfnamefont {P.}~\bibnamefont {Lodahl}},\ }\bibfield  {title} {\bibinfo {title} {Deterministic photon source interfaced with a programmable silicon-nitride integrated circuit},\ }\href
  {https://doi.org/10.1038/s41534-023-00761-1} {\bibfield  {journal} {\bibinfo  {journal} {npj Quantum Information}\ }\textbf {\bibinfo {volume} {9}},\ \bibinfo {pages} {94} (\bibinfo {year} {2023})}\BibitemShut {NoStop}%
\bibitem [{\citenamefont {Chanana}\ \emph {et~al.}(2022)\citenamefont {Chanana}, \citenamefont {Larocque}, \citenamefont {Moreira}, \citenamefont {Carolan}, \citenamefont {Guha}, \citenamefont {Melo}, \citenamefont {Anant}, \citenamefont {Song}, \citenamefont {Englund}, \citenamefont {Blumenthal}, \citenamefont {Srinivasan},\ and\ \citenamefont {Davanco}}]{SourceIntegrated}%
  \BibitemOpen
  \bibfield  {author} {\bibinfo {author} {\bibfnamefont {A.}~\bibnamefont {Chanana}}, \bibinfo {author} {\bibfnamefont {H.}~\bibnamefont {Larocque}}, \bibinfo {author} {\bibfnamefont {R.}~\bibnamefont {Moreira}}, \bibinfo {author} {\bibfnamefont {J.}~\bibnamefont {Carolan}}, \bibinfo {author} {\bibfnamefont {B.}~\bibnamefont {Guha}}, \bibinfo {author} {\bibfnamefont {E.~G.}\ \bibnamefont {Melo}}, \bibinfo {author} {\bibfnamefont {V.}~\bibnamefont {Anant}}, \bibinfo {author} {\bibfnamefont {J.}~\bibnamefont {Song}}, \bibinfo {author} {\bibfnamefont {D.}~\bibnamefont {Englund}}, \bibinfo {author} {\bibfnamefont {D.~J.}\ \bibnamefont {Blumenthal}}, \bibinfo {author} {\bibfnamefont {K.}~\bibnamefont {Srinivasan}},\ and\ \bibinfo {author} {\bibfnamefont {M.}~\bibnamefont {Davanco}},\ }\bibfield  {title} {\bibinfo {title} {Ultra-low loss quantum photonic circuits integrated with single quantum emitters},\ }\href {https://doi.org/10.1038/s41467-022-35332-z} {\bibfield  {journal} {\bibinfo  {journal} {Nature
  Communications}\ }\textbf {\bibinfo {volume} {13}},\ \bibinfo {pages} {7693} (\bibinfo {year} {2022})}\BibitemShut {NoStop}%
\bibitem [{\citenamefont {Ferrari}\ \emph {et~al.}(2018)\citenamefont {Ferrari}, \citenamefont {Schuck},\ and\ \citenamefont {Pernice}}]{IntegratedDetectors}%
  \BibitemOpen
  \bibfield  {author} {\bibinfo {author} {\bibfnamefont {S.}~\bibnamefont {Ferrari}}, \bibinfo {author} {\bibfnamefont {C.}~\bibnamefont {Schuck}},\ and\ \bibinfo {author} {\bibfnamefont {W.}~\bibnamefont {Pernice}},\ }\bibfield  {title} {\bibinfo {title} {Waveguide-integrated superconducting nanowire single-photon detectors},\ }\href {https://doi.org/doi:10.1515/nanoph-2018-0059} {\bibfield  {journal} {\bibinfo  {journal} {Nanophotonics}\ }\textbf {\bibinfo {volume} {7}},\ \bibinfo {pages} {1725} (\bibinfo {year} {2018})}\BibitemShut {NoStop}%
\bibitem [{\citenamefont {Beutel}\ \emph {et~al.}(2022)\citenamefont {Beutel}, \citenamefont {Grottke}, \citenamefont {Wolff}, \citenamefont {Schuck},\ and\ \citenamefont {Pernice}}]{CryoPS}%
  \BibitemOpen
  \bibfield  {author} {\bibinfo {author} {\bibfnamefont {F.}~\bibnamefont {Beutel}}, \bibinfo {author} {\bibfnamefont {T.}~\bibnamefont {Grottke}}, \bibinfo {author} {\bibfnamefont {M.~A.}\ \bibnamefont {Wolff}}, \bibinfo {author} {\bibfnamefont {C.}~\bibnamefont {Schuck}},\ and\ \bibinfo {author} {\bibfnamefont {W.~H.~P.}\ \bibnamefont {Pernice}},\ }\bibfield  {title} {\bibinfo {title} {Cryo-compatible opto-mechanical low-voltage phase-modulator integrated with superconducting single-photon detectors},\ }\href {https://doi.org/10.1364/OE.462163} {\bibfield  {journal} {\bibinfo  {journal} {Opt. Express}\ }\textbf {\bibinfo {volume} {30}},\ \bibinfo {pages} {30066} (\bibinfo {year} {2022})}\BibitemShut {NoStop}%
\bibitem [{\citenamefont {McCaw}\ \emph {et~al.}(2024)\citenamefont {McCaw}, \citenamefont {Ewaniuk}, \citenamefont {Shastri},\ and\ \citenamefont {Rotenberg}}]{QDphaseshifts}%
  \BibitemOpen
  \bibfield  {author} {\bibinfo {author} {\bibfnamefont {A.}~\bibnamefont {McCaw}}, \bibinfo {author} {\bibfnamefont {J.}~\bibnamefont {Ewaniuk}}, \bibinfo {author} {\bibfnamefont {B.~J.}\ \bibnamefont {Shastri}},\ and\ \bibinfo {author} {\bibfnamefont {N.}~\bibnamefont {Rotenberg}},\ }\bibfield  {title} {\bibinfo {title} {Reconfigurable quantum photonic circuits based on quantum dots},\ }\href {https://doi.org/doi:10.1515/nanoph-2024-0044} {\bibfield  {journal} {\bibinfo  {journal} {Nanophotonics}\ }\textbf {\bibinfo {volume} {13}},\ \bibinfo {pages} {2951} (\bibinfo {year} {2024})}\BibitemShut {NoStop}%
\bibitem [{\citenamefont {Bell}\ and\ \citenamefont {Walmsley}(2021)}]{Bell}%
  \BibitemOpen
  \bibfield  {author} {\bibinfo {author} {\bibfnamefont {B.~A.}\ \bibnamefont {Bell}}\ and\ \bibinfo {author} {\bibfnamefont {I.~A.}\ \bibnamefont {Walmsley}},\ }\bibfield  {title} {\bibinfo {title} {Further compactifying linear optical unitaries},\ }\href {https://doi.org/10.1063/5.0053421} {\bibfield  {journal} {\bibinfo  {journal} {APL Photonics}\ }\textbf {\bibinfo {volume} {6}},\ \bibinfo {pages} {070804} (\bibinfo {year} {2021})},\ \Eprint {https://arxiv.org/abs/https://pubs.aip.org/aip/app/article-pdf/doi/10.1063/5.0053421/14079379/070804\_1\_online.pdf} {https://pubs.aip.org/aip/app/article-pdf/doi/10.1063/5.0053421/14079379/070804\_1\_online.pdf} \BibitemShut {NoStop}%
\bibitem [{\citenamefont {McClean}\ \emph {et~al.}(2018)\citenamefont {McClean}, \citenamefont {Boixo}, \citenamefont {Smelyanskiy}, \citenamefont {Babbush},\ and\ \citenamefont {Neven}}]{BarrenPlateaus}%
  \BibitemOpen
  \bibfield  {author} {\bibinfo {author} {\bibfnamefont {J.~R.}\ \bibnamefont {McClean}}, \bibinfo {author} {\bibfnamefont {S.}~\bibnamefont {Boixo}}, \bibinfo {author} {\bibfnamefont {V.~N.}\ \bibnamefont {Smelyanskiy}}, \bibinfo {author} {\bibfnamefont {R.}~\bibnamefont {Babbush}},\ and\ \bibinfo {author} {\bibfnamefont {H.}~\bibnamefont {Neven}},\ }\bibfield  {title} {\bibinfo {title} {Barren plateaus in quantum neural network training landscapes},\ }\href {https://doi.org/10.1038/s41467-018-07090-4} {\bibfield  {journal} {\bibinfo  {journal} {Nature Communications}\ }\textbf {\bibinfo {volume} {9}},\ \bibinfo {pages} {4812} (\bibinfo {year} {2018})}\BibitemShut {NoStop}%
\bibitem [{\citenamefont {Larocca}\ \emph {et~al.}(2024)\citenamefont {Larocca}, \citenamefont {Thanasilp}, \citenamefont {Wang}, \citenamefont {Sharma}, \citenamefont {Biamonte}, \citenamefont {Coles}, \citenamefont {Cincio}, \citenamefont {McClean}, \citenamefont {Holmes},\ and\ \citenamefont {Cerezo}}]{BarrenPlateausReview}%
  \BibitemOpen
  \bibfield  {author} {\bibinfo {author} {\bibfnamefont {M.}~\bibnamefont {Larocca}}, \bibinfo {author} {\bibfnamefont {S.}~\bibnamefont {Thanasilp}}, \bibinfo {author} {\bibfnamefont {S.}~\bibnamefont {Wang}}, \bibinfo {author} {\bibfnamefont {K.}~\bibnamefont {Sharma}}, \bibinfo {author} {\bibfnamefont {J.}~\bibnamefont {Biamonte}}, \bibinfo {author} {\bibfnamefont {P.~J.}\ \bibnamefont {Coles}}, \bibinfo {author} {\bibfnamefont {L.}~\bibnamefont {Cincio}}, \bibinfo {author} {\bibfnamefont {J.~R.}\ \bibnamefont {McClean}}, \bibinfo {author} {\bibfnamefont {Z.}~\bibnamefont {Holmes}},\ and\ \bibinfo {author} {\bibfnamefont {M.}~\bibnamefont {Cerezo}},\ }\href@noop {} {\bibinfo {title} {A review of barren plateaus in variational quantum computing}} (\bibinfo {year} {2024}),\ \Eprint {https://arxiv.org/abs/2405.00781} {arXiv:2405.00781 [quant-ph]} \BibitemShut {NoStop}%
\bibitem [{\citenamefont {{Kitaev}}(1997)}]{Kitaev}%
  \BibitemOpen
  \bibfield  {author} {\bibinfo {author} {\bibfnamefont {A.~Y.}\ \bibnamefont {{Kitaev}}},\ }\bibfield  {title} {\bibinfo {title} {{Quantum computations: algorithms and error correction}},\ }\href {https://doi.org/10.1070/RM1997v052n06ABEH002155} {\bibfield  {journal} {\bibinfo  {journal} {Russian Mathematical Surveys}\ }\textbf {\bibinfo {volume} {52}},\ \bibinfo {pages} {1191} (\bibinfo {year} {1997})}\BibitemShut {NoStop}%
\bibitem [{\citenamefont {Shor}(1995)}]{shor1995scheme}%
  \BibitemOpen
  \bibfield  {author} {\bibinfo {author} {\bibfnamefont {P.~W.}\ \bibnamefont {Shor}},\ }\bibfield  {title} {\bibinfo {title} {Scheme for reducing decoherence in quantum computer memory},\ }\href@noop {} {\bibfield  {journal} {\bibinfo  {journal} {Physical review A}\ }\textbf {\bibinfo {volume} {52}},\ \bibinfo {pages} {R2493} (\bibinfo {year} {1995})}\BibitemShut {NoStop}%
\bibitem [{\citenamefont {Ekert}(1991)}]{ekert1991quantum}%
  \BibitemOpen
  \bibfield  {author} {\bibinfo {author} {\bibfnamefont {A.~K.}\ \bibnamefont {Ekert}},\ }\bibfield  {title} {\bibinfo {title} {Quantum cryptography based on bell’s theorem},\ }\href@noop {} {\bibfield  {journal} {\bibinfo  {journal} {Physical review letters}\ }\textbf {\bibinfo {volume} {67}},\ \bibinfo {pages} {661} (\bibinfo {year} {1991})}\BibitemShut {NoStop}%
\bibitem [{\citenamefont {Bartolucci}\ \emph {et~al.}(2021)\citenamefont {Bartolucci}, \citenamefont {Birchall}, \citenamefont {Gimeno-Segovia}, \citenamefont {Johnston}, \citenamefont {Kieling}, \citenamefont {Pant}, \citenamefont {Rudolph}, \citenamefont {Smith}, \citenamefont {Sparrow},\ and\ \citenamefont {Vidrighin}}]{bartolucci2021creation}%
  \BibitemOpen
  \bibfield  {author} {\bibinfo {author} {\bibfnamefont {S.}~\bibnamefont {Bartolucci}}, \bibinfo {author} {\bibfnamefont {P.~M.}\ \bibnamefont {Birchall}}, \bibinfo {author} {\bibfnamefont {M.}~\bibnamefont {Gimeno-Segovia}}, \bibinfo {author} {\bibfnamefont {E.}~\bibnamefont {Johnston}}, \bibinfo {author} {\bibfnamefont {K.}~\bibnamefont {Kieling}}, \bibinfo {author} {\bibfnamefont {M.}~\bibnamefont {Pant}}, \bibinfo {author} {\bibfnamefont {T.}~\bibnamefont {Rudolph}}, \bibinfo {author} {\bibfnamefont {J.}~\bibnamefont {Smith}}, \bibinfo {author} {\bibfnamefont {C.}~\bibnamefont {Sparrow}},\ and\ \bibinfo {author} {\bibfnamefont {M.~D.}\ \bibnamefont {Vidrighin}},\ }\href@noop {} {\bibinfo {title} {Creation of entangled photonic states using linear optics}} (\bibinfo {year} {2021}),\ \Eprint {https://arxiv.org/abs/2106.13825} {arXiv:2106.13825 [quant-ph]} \BibitemShut {NoStop}%
\bibitem [{\citenamefont {Fldzhyan}\ \emph {et~al.}(2023)\citenamefont {Fldzhyan}, \citenamefont {Saygin},\ and\ \citenamefont {Kulik}}]{Programmable2qubits}%
  \BibitemOpen
  \bibfield  {author} {\bibinfo {author} {\bibfnamefont {S.~A.}\ \bibnamefont {Fldzhyan}}, \bibinfo {author} {\bibfnamefont {M.~Y.}\ \bibnamefont {Saygin}},\ and\ \bibinfo {author} {\bibfnamefont {S.~P.}\ \bibnamefont {Kulik}},\ }\bibfield  {title} {\bibinfo {title} {Programmable heralded linear optical generation of two-qubit states},\ }\href {https://doi.org/10.1103/PhysRevApplied.20.054030} {\bibfield  {journal} {\bibinfo  {journal} {Phys. Rev. Appl.}\ }\textbf {\bibinfo {volume} {20}},\ \bibinfo {pages} {054030} (\bibinfo {year} {2023})}\BibitemShut {NoStop}%
\bibitem [{\citenamefont {Muller}(1959)}]{RandomPointsnSphere}%
  \BibitemOpen
  \bibfield  {author} {\bibinfo {author} {\bibfnamefont {M.~E.}\ \bibnamefont {Muller}},\ }\bibfield  {title} {\bibinfo {title} {A note on a method for generating points uniformly on n-dimensional spheres},\ }\href {https://doi.org/10.1145/377939.377946} {\bibfield  {journal} {\bibinfo  {journal} {Commun. ACM}\ }\textbf {\bibinfo {volume} {2}},\ \bibinfo {pages} {19–20} (\bibinfo {year} {1959})}\BibitemShut {NoStop}%
\bibitem [{\citenamefont {Calsamiglia}\ and\ \citenamefont {L{\"u}tkenhaus}(2001)}]{LO_BELL_DISCRIMINATOR}%
  \BibitemOpen
  \bibfield  {author} {\bibinfo {author} {\bibfnamefont {J.}~\bibnamefont {Calsamiglia}}\ and\ \bibinfo {author} {\bibfnamefont {N.}~\bibnamefont {L{\"u}tkenhaus}},\ }\bibfield  {title} {\bibinfo {title} {Maximum efficiency of a linear-optical bell-state analyzer},\ }\href {https://doi.org/10.1007/s003400000484} {\bibfield  {journal} {\bibinfo  {journal} {Applied Physics B}\ }\textbf {\bibinfo {volume} {72}},\ \bibinfo {pages} {67} (\bibinfo {year} {2001})}\BibitemShut {NoStop}%
\bibitem [{\citenamefont {Yamazaki}\ \emph {et~al.}(2023)\citenamefont {Yamazaki}, \citenamefont {Ikuta},\ and\ \citenamefont {Yamamoto}}]{Busted_discriminator}%
  \BibitemOpen
  \bibfield  {author} {\bibinfo {author} {\bibfnamefont {T.}~\bibnamefont {Yamazaki}}, \bibinfo {author} {\bibfnamefont {R.}~\bibnamefont {Ikuta}},\ and\ \bibinfo {author} {\bibfnamefont {T.}~\bibnamefont {Yamamoto}},\ }\href@noop {} {\bibinfo {title} {Stabilizer formalism in linear optics and application to bell-state discrimination}} (\bibinfo {year} {2023}),\ \Eprint {https://arxiv.org/abs/2301.06551} {arXiv:2301.06551 [quant-ph]} \BibitemShut {NoStop}%
\bibitem [{\citenamefont {Zhong}\ and\ \citenamefont {Jiang}(2002)}]{zhong2002theoretical}%
  \BibitemOpen
  \bibfield  {author} {\bibinfo {author} {\bibfnamefont {C.}~\bibnamefont {Zhong}}\ and\ \bibinfo {author} {\bibfnamefont {Q.}~\bibnamefont {Jiang}},\ }\bibfield  {title} {\bibinfo {title} {Theoretical study on perpendicular magnetoelectric coupling in ferroelectromagnet system},\ }\href {https://doi.org/https://doi.org/10.1016/S0038-1098(02)00215-6} {\bibfield  {journal} {\bibinfo  {journal} {Solid State Communications}\ }\textbf {\bibinfo {volume} {122}},\ \bibinfo {pages} {601} (\bibinfo {year} {2002})}\BibitemShut {NoStop}%
\bibitem [{\citenamefont {Banchi}\ \emph {et~al.}(2017)\citenamefont {Banchi}, \citenamefont {Coutinho}, \citenamefont {Godsil},\ and\ \citenamefont {Severini}}]{banchi2017pretty}%
  \BibitemOpen
  \bibfield  {author} {\bibinfo {author} {\bibfnamefont {L.}~\bibnamefont {Banchi}}, \bibinfo {author} {\bibfnamefont {G.}~\bibnamefont {Coutinho}}, \bibinfo {author} {\bibfnamefont {C.}~\bibnamefont {Godsil}},\ and\ \bibinfo {author} {\bibfnamefont {S.}~\bibnamefont {Severini}},\ }\bibfield  {title} {\bibinfo {title} {{Pretty good state transfer in qubit chains—The Heisenberg Hamiltonian}},\ }\href {https://doi.org/10.1063/1.4978327} {\bibfield  {journal} {\bibinfo  {journal} {Journal of Mathematical Physics}\ }\textbf {\bibinfo {volume} {58}},\ \bibinfo {pages} {032202} (\bibinfo {year} {2017})}\BibitemShut {NoStop}%
\bibitem [{\citenamefont {Weihong}\ \emph {et~al.}(1999)\citenamefont {Weihong}, \citenamefont {McKenzie},\ and\ \citenamefont {Singh}}]{weihong1999phase}%
  \BibitemOpen
  \bibfield  {author} {\bibinfo {author} {\bibfnamefont {Z.}~\bibnamefont {Weihong}}, \bibinfo {author} {\bibfnamefont {R.~H.}\ \bibnamefont {McKenzie}},\ and\ \bibinfo {author} {\bibfnamefont {R.~R.~P.}\ \bibnamefont {Singh}},\ }\bibfield  {title} {\bibinfo {title} {Phase diagram for a class of spin-$\frac{1}{2}$ heisenberg models interpolating between the square-lattice, the triangular-lattice, and the linear-chain limits},\ }\href {https://doi.org/10.1103/PhysRevB.59.14367} {\bibfield  {journal} {\bibinfo  {journal} {Phys. Rev. B}\ }\textbf {\bibinfo {volume} {59}},\ \bibinfo {pages} {14367} (\bibinfo {year} {1999})}\BibitemShut {NoStop}%
\bibitem [{\citenamefont {Zheng}\ \emph {et~al.}(2006)\citenamefont {Zheng}, \citenamefont {Fj\ae{}restad}, \citenamefont {Singh}, \citenamefont {McKenzie},\ and\ \citenamefont {Coldea}}]{zheng2006excitation}%
  \BibitemOpen
  \bibfield  {author} {\bibinfo {author} {\bibfnamefont {W.}~\bibnamefont {Zheng}}, \bibinfo {author} {\bibfnamefont {J.~O.}\ \bibnamefont {Fj\ae{}restad}}, \bibinfo {author} {\bibfnamefont {R.~R.~P.}\ \bibnamefont {Singh}}, \bibinfo {author} {\bibfnamefont {R.~H.}\ \bibnamefont {McKenzie}},\ and\ \bibinfo {author} {\bibfnamefont {R.}~\bibnamefont {Coldea}},\ }\bibfield  {title} {\bibinfo {title} {Excitation spectra of the spin-$\frac{1}{2}$ triangular-lattice heisenberg antiferromagnet},\ }\href {https://doi.org/10.1103/PhysRevB.74.224420} {\bibfield  {journal} {\bibinfo  {journal} {Phys. Rev. B}\ }\textbf {\bibinfo {volume} {74}},\ \bibinfo {pages} {224420} (\bibinfo {year} {2006})}\BibitemShut {NoStop}%
\bibitem [{\citenamefont {Killoran}\ \emph {et~al.}(2019)\citenamefont {Killoran}, \citenamefont {Bromley}, \citenamefont {Arrazola}, \citenamefont {Schuld}, \citenamefont {Quesada},\ and\ \citenamefont {Lloyd}}]{CV_NN}%
  \BibitemOpen
  \bibfield  {author} {\bibinfo {author} {\bibfnamefont {N.}~\bibnamefont {Killoran}}, \bibinfo {author} {\bibfnamefont {T.~R.}\ \bibnamefont {Bromley}}, \bibinfo {author} {\bibfnamefont {J.~M.}\ \bibnamefont {Arrazola}}, \bibinfo {author} {\bibfnamefont {M.}~\bibnamefont {Schuld}}, \bibinfo {author} {\bibfnamefont {N.}~\bibnamefont {Quesada}},\ and\ \bibinfo {author} {\bibfnamefont {S.}~\bibnamefont {Lloyd}},\ }\bibfield  {title} {\bibinfo {title} {Continuous-variable quantum neural networks},\ }\href {https://doi.org/10.1103/PhysRevResearch.1.033063} {\bibfield  {journal} {\bibinfo  {journal} {Phys. Rev. Res.}\ }\textbf {\bibinfo {volume} {1}},\ \bibinfo {pages} {033063} (\bibinfo {year} {2019})}\BibitemShut {NoStop}%
\bibitem [{\citenamefont {Liu}\ \emph {et~al.}(2024)\citenamefont {Liu}, \citenamefont {Wang}, \citenamefont {Vaidya}, \citenamefont {Ruehle}, \citenamefont {Halverson}, \citenamefont {Soljačić}, \citenamefont {Hou},\ and\ \citenamefont {Tegmark}}]{KAN_NN}%
  \BibitemOpen
  \bibfield  {author} {\bibinfo {author} {\bibfnamefont {Z.}~\bibnamefont {Liu}}, \bibinfo {author} {\bibfnamefont {Y.}~\bibnamefont {Wang}}, \bibinfo {author} {\bibfnamefont {S.}~\bibnamefont {Vaidya}}, \bibinfo {author} {\bibfnamefont {F.}~\bibnamefont {Ruehle}}, \bibinfo {author} {\bibfnamefont {J.}~\bibnamefont {Halverson}}, \bibinfo {author} {\bibfnamefont {M.}~\bibnamefont {Soljačić}}, \bibinfo {author} {\bibfnamefont {T.~Y.}\ \bibnamefont {Hou}},\ and\ \bibinfo {author} {\bibfnamefont {M.}~\bibnamefont {Tegmark}},\ }\href@noop {} {\bibinfo {title} {Kan: Kolmogorov-arnold networks}} (\bibinfo {year} {2024}),\ \Eprint {https://arxiv.org/abs/2404.19756} {arXiv:2404.19756 [cs.LG]} \BibitemShut {NoStop}%
\bibitem [{\citenamefont {Kaelo}\ and\ \citenamefont {Ali}(2006)}]{kaelo2006some}%
  \BibitemOpen
  \bibfield  {author} {\bibinfo {author} {\bibfnamefont {P.}~\bibnamefont {Kaelo}}\ and\ \bibinfo {author} {\bibfnamefont {M.}~\bibnamefont {Ali}},\ }\bibfield  {title} {\bibinfo {title} {Some variants of the controlled random search algorithm for global optimization},\ }\href {https://doi.org/10.1007/s10957-006-9101-0} {\bibfield  {journal} {\bibinfo  {journal} {Journal of optimization theory and applications}\ }\textbf {\bibinfo {volume} {130}},\ \bibinfo {pages} {253} (\bibinfo {year} {2006})}\BibitemShut {NoStop}%
\bibitem [{\citenamefont {Powell}\ \emph {et~al.}(2009)\citenamefont {Powell} \emph {et~al.}}]{powell2009bobyqa}%
  \BibitemOpen
  \bibfield  {author} {\bibinfo {author} {\bibfnamefont {M.~J.}\ \bibnamefont {Powell}} \emph {et~al.},\ }\bibfield  {title} {\bibinfo {title} {The bobyqa algorithm for bound constrained optimization without derivatives},\ }\href@noop {} {\bibfield  {journal} {\bibinfo  {journal} {Cambridge NA Report NA2009/06, University of Cambridge, Cambridge}\ }\textbf {\bibinfo {volume} {26}},\ \bibinfo {pages} {26} (\bibinfo {year} {2009})}\BibitemShut {NoStop}%
\bibitem [{\citenamefont {Johnson}\ \emph {et~al.}(2014)\citenamefont {Johnson} \emph {et~al.}}]{johnson2014nlopt}%
  \BibitemOpen
  \bibfield  {author} {\bibinfo {author} {\bibfnamefont {S.~G.}\ \bibnamefont {Johnson}} \emph {et~al.},\ }\bibfield  {title} {\bibinfo {title} {The nlopt nonlinear-optimization package, 2014},\ }\href {http://ab-initio.mit.edu/nlopt} {\bibfield  {journal} {\bibinfo  {journal} {URL http://ab-initio.mit.edu/nlopt}\ } (\bibinfo {year} {2014})}\BibitemShut {NoStop}%
\end{thebibliography}%

\end{document}